\documentclass[tighten, twocolumn, twocolappendix]{aastex631}

\usepackage{amsmath}
\usepackage{tabularx} 

\begin{document}

\title{GRMHD Study of Accretion onto time-like Naked Singularities}

\author[0000-0001-8213-646X]{Akhil Uniyal}
\affiliation{Tsung-Dao Lee Institute, Shanghai Jiao Tong University, Shengrong Road 520, Shanghai, 201210, People’s Republic of China}

\author[0000-0002-4064-0446]{Indu K. Dihingia}
\affiliation{Tsung-Dao Lee Institute, Shanghai Jiao Tong University, Shengrong Road 520, Shanghai, 201210, People’s Republic of China}

\author[0000-0002-8131-6730]{Yosuke Mizuno}
\affiliation{Tsung-Dao Lee Institute, Shanghai Jiao Tong University, Shengrong Road 520, Shanghai, 201210, People’s Republic of China}
\affiliation{School of Physics and Astronomy, Shanghai Jiao Tong University, 800 Dongchuan Road, Shanghai, 200240, People’s Republic of
China}
\affiliation{Key Laboratory for Particle Physics, Astrophysics and Cosmology (MOE), Shanghai Key Laboratory for Particle Physics and Cosmology, Shanghai Jiao-Tong University,800 Dongchuan Road, Shanghai, 200240, People's Republic of China}
\affiliation{Institut f\"{u}r Theoretische Physik, Goethe-Universit\"{a}t Frankfurt, Max-von-Laue-Strasse 1, D-60438 Frankfurt am Main, Germany}

\author[0000-0001-9043-8062]{W{\l}odek Klu\'{z}niak}
\affiliation{Nicolaus Copernicus Astronomical Center of the Polish Academy of Sciences, Bartycka 18, 00-716 Warszawa, Poland}

\begin{abstract}
Naked singularities (NkS) are solutions to the Einstein field equations that violate the cosmic censorship conjecture. Recent studies indicate that these objects may serve as compelling mimickers of black hole shadows. In this work, we investigate the accretion dynamics of selected time-like naked singularities using general relativistic magnetohydrodynamic (GRMHD) simulations. Our objective is to determine whether naked singularities exhibit distinct signatures compared to black holes. We find that, unlike black holes, naked singularities exhibit a centrifugal barrier that prevents direct accretion of rotating matter onto the NkS. Despite reduced magnetization in the funnel region, these objects are capable of generating jet powers comparable to those observed in black holes. Additionally, we observe that accreting matter releases gravitational energy as it is driven towards the NkS, powering the strong outflow via local fluid pressure gradient or magnetic pressure forces.
\end{abstract}


\section{Introduction} \label{sec:intro}

The solution of Einstein's General Relativity (GR) suggests the existence of compact objects such as black holes (BHs), the naked singularity (NkS), and wormholes (WHs) \citep{Weinberg:1972kfs, Wald:1984rg, Chandrasekhar:1985kt, Hartle:2003yu}. The existence of the NkS would violate the strong cosmic censorship conjecture, according to which singularities should be hidden by the event horizons \citep{Penrose:1969pc}. 
However, there is no observational verification for such a strong assertion, and in recent years, it has been established that NkS and BHs are both possible outcomes of the gravitational collapse of a massive matter cloud in 
various scenarios, such as collapse of gravitating dust~\citep{Eardley:1978tr, Christodoulou:1984mz, Singh:1994tb}, self-similar collapse \citep{Ori:1987hg, Ori:1989ps, Foglizzo:1993unt}, scalar fields \citep{Christodoulou:1994hg, Giambo:2005se, Bhattacharya2010COLLAPSEAD}, perfect fluids, and other matter fields \citep{Harada:1998wb, Harada:1999jf, Goswami:2002ds, Giambo:2004xh, VillasdaRocha:2000vi}, including fluid at non-zero pressures \citep{joshi2007gravitational, Joshi:2011zm}. Further study suggests that the spherical gravitational collapse can also lead to the NkS as an end-stage in a cosmological scenario \citep{Bhattacharya:2017chr}.
Analytical calculations suggests that NkS can also cast a similar shadow as in the case of the BHs \citep[e.g.,][]{Shaikh:2018lcc, Joshi:2020tlq, Dey:2020bgo}. NkS remains a promising candidate in BH mimickers for creating ring-like shadow images observed by the Event Horizon Telescope \citep{EventHorizonTelescope:2019dse, EventHorizonTelescope:2022wkp}. 
Therefore, the study of naked singularities provides valuable insights into the gravitational collapse process and plays a crucial role in testing the cosmic censorship conjecture. 

Several other analytical studies have examined the properties of accretion disks, gravitational lensing, and other phenomena to differentiate between BHs and naked singularities \citep{Gyulchev:2008ff, Kovacs:2010xm}. However, a full radiative General Relativistic Magnetohydrodynamics (GRMHD) simulation is still missing to analyze the realistic scenario. Recent studies have made progress in this direction, utilizing analytic GR calculations, as well as simulations: radiative hydrodynamic, GRMHD, and resistive MHD, to distinguish Reissner-Nordström (RN) and Kerr-type naked singularities from BHs \citep{ Mishra:2024bpl, Kluzniak:2024cxm, Dihingia:2024cch, Cemeljic:2025bqz}. This work focuses on the extension of these studies to other metrics with the help of full GRMHD simulations.

In this work, we perform
$3D$ GRMHD simulations around three different types of rapidly spinning NkS: $(a)$ Kerr NkS with spin $a>1$, also known as the superspinar (SS); $(b)$ Joshi-Malafarina-Narayan (JMN) \citep{Joshi:2011zm}; and $(c)$ Janis-Newman-Winicour (JNW) \citep{Janis:1968zz}. To perform the GRMHD simulations, we use the horizon-penetrating (HP) form of a general stationary and axisymmetric metric known as Azreg-A{\"i}nou (AA) metric \citep{Azreg-Ainou:2014aqa, Azreg-Ainou:2014pra} developed by \cite{Kocherlakota:2023vff}.
The simulations are performed using the GRMHD code \texttt{BHAC} \citep{Porth:2016rfi, Olivares:2019dsc}. The length and time are expressed in units of $r_g=GM/c^2$ and $t_g=r_g/c$, respectively;
$M$ is the mass of the NkS.

The outline of the paper is as follows: In Sec.~\ref{sec:2}, we introduce the NkS space-times that are used in this work. In Sec.~\ref{sec:3}, we describe the setup for the initial conditions to perform the GRMHD simulation. In Sec.~\ref{sec:4}, we show the results obtained in 3D simulations and compare these compact objects. In Sec.~\ref{sec:5} we discuss our findings, and finally in Sec.~\ref{sec:6}, we conclude the outcome of the study. 


\section{Naked-Singularity} \label{sec:2}

Although we discuss naked singularities and not black holes, for convenience, we use the HP form of the metric for any general axisymmetric space-time \citep{Kocherlakota:2023vff}, in geometrized units (the speed of light, $c = 1$, the gravitational constant, $G = 1$):
\begin{align} \label{eq:AA_siKS_C}
\mathrm{d}s^2 =&\ \left[-\left(1-\frac{2F}{\Sigma}\right)\mathrm{d}\tau^2 +  \left(1+\frac{2F}{\Sigma}\right)\mathrm{d}r^2 \right. \nonumber \\
&\ \left. + \Sigma~\mathrm{d}\vartheta^2  + \frac{\Pi}{\Sigma}\sin^2{\vartheta}~\mathrm{d}\phi^2 
- 2\frac{2F}{\Sigma}a\sin^2{\vartheta}~\mathrm{d}\tau\mathrm{d}\phi \right. \nonumber \\
&\ \left. + 2\frac{2F}{\Sigma}\mathrm{d}\tau\mathrm{d}r - 2\left(1+\frac{2F}{\Sigma}\right)a\sin^2{\vartheta}~\mathrm{d}r\mathrm{d}\phi\right]\,,
\end{align}
where,
\begin{align} \label{eq:AA_Staionary_Metric_Functions_fgh}
2F(r) =&\ R^2/\sqrt{g}-(f/g)R^2\,, \\ 
\Delta(r) =&\ (f/g)R^2 + a^2\,, \nonumber \\
\Sigma(r, \theta) =&\ R^2/\sqrt{g} + a^2\cos^2{\theta}\,,\nonumber \\ 
\Pi(r, \theta) =&\ (R^2/\sqrt{g}+a^2)^2 - \Delta a^2\sin^2{\theta}\,. \nonumber 
\end{align}
Here $a$ describes the spin of the spacetime and $f$ and $g$ are the metric components of the corresponding static and spherically symmetric metric $(-f, g/f, R^2, R^2 \sin^2{\theta})$.

\subsection{Kerr metric}\label{kerr-type}

Kerr space-time can be written in terms of the functions used in the HP form of the AA metric as:
\begin{align}
  f(r) =&\ 1 - \frac{2M}{r}, \\
  g(r) = &\ 1, \\
  R^2(r) =&\ r^2,
\end{align}
where we have a Kerr-metric NkS for $a/M>1$.
In this familiar case, $R=r$.

\subsection{JMN}\label{jmn}

In this work, we also consider Joshi-Malafarina-Narayan (JMN) \citep{Joshi:2011zm} space-time. It can be written in terms of the functions used in the HP form of the AA metric as:
\begin{align}
 f(r) =&\ (1-M_0)\left(\frac{R}{R_b}\right)^{M_0/(1-M_0)}, \\
 g(r) = &\ \frac{f(r)}{1-M_0}, \\
  R^2(r) =&\ r^2,
\end{align}
where $R_b$ is the matching radius, $M_0=2/R_b$.

\subsection{JNW}\label{jnw}

Another NkS space-time we considered is Janis-Newman-Winicour (JNW) space-time \citep{Janis:1968zz}. The solution was first discovered by \cite{Fisher:1948yn} and also \cite{Buchdahl:1959zz}. Later, Janis, Newman, and Winicour rediscovered the solution in an isotropic coordinate system. Thereafter, \cite{Wyman:1981bd} discovered once again the solution in the Schwarzschild coordinate system. It can be written in terms of the functions used in the HP form of the AA metric as:
\begin{align}
  f(r) =&\ \left(1-\frac{r_\star}{r}\right)^{1-\hat{\nu}}, \\
  g(r) = &\ 1, \\
  R^2(r) =&\ r^2\left(1-\frac{r_\star}{r}\right)^{\hat{\nu}},
\end{align}
where $r_\star=2/(1-\hat{\nu})$ is the curvature singularity radius.
\subsection{Effective potential}\label{effpot}
\begin{figure}[t]
    \centering
  \includegraphics[width=0.5\textwidth]{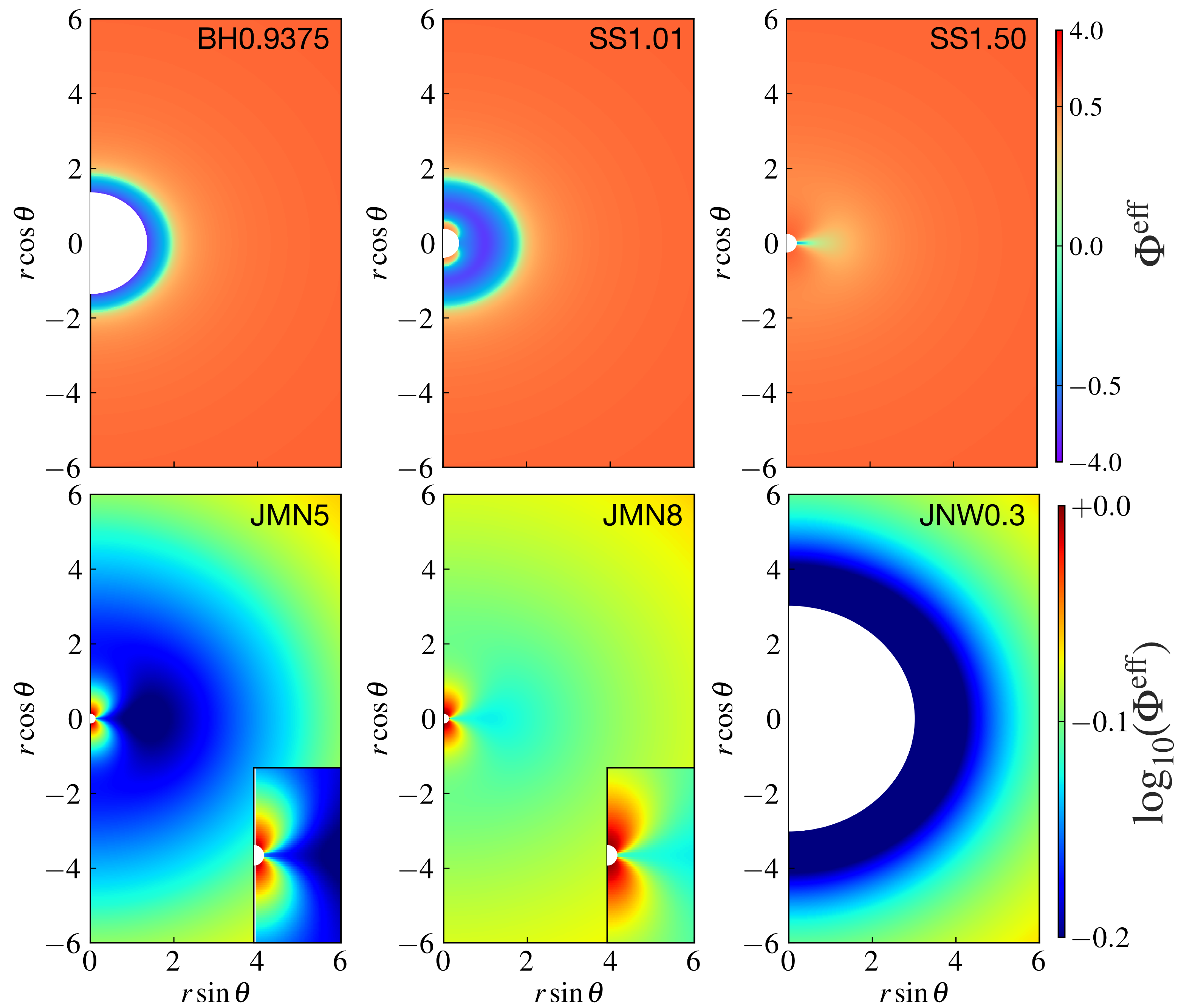}
    \caption{Effective potential in the poloidal plane for zero specific angular momentum, $\lambda=0$. A Kerr black hole and five different naked singularities are shown.}
    \label{pot}
\end{figure}

For all three NkS space-times, we perform two cases for each. In the superspinar, we consider dimensionless spin $a_*\equiv a/M=1.01$ (SS1.01) and $a_*=1.50$ (SS1.50). For the other metrics, we fix the spin at $a_*=0.9375$ and vary the other space-time related parameters. In the case of JMN space-time, we consider two models with different matching radii $R_b=5$ (JMN5)
and $R_b=8$ (JMN8). Similarly, for JNW space-time, we consider two models with different scalar charges $\hat{\nu}=0.3$ (JNW0.3) and $\hat{\nu}=0.6$ (JNW0.6). The JMN and JNW singularities are point-like and spherical, respectively. The Kerr metric singularity is not spherically symmetric \citep{Chandrasekhar:1985kt}.

\begin{figure}[t]
    \centering
  \includegraphics[width=0.44\textwidth]{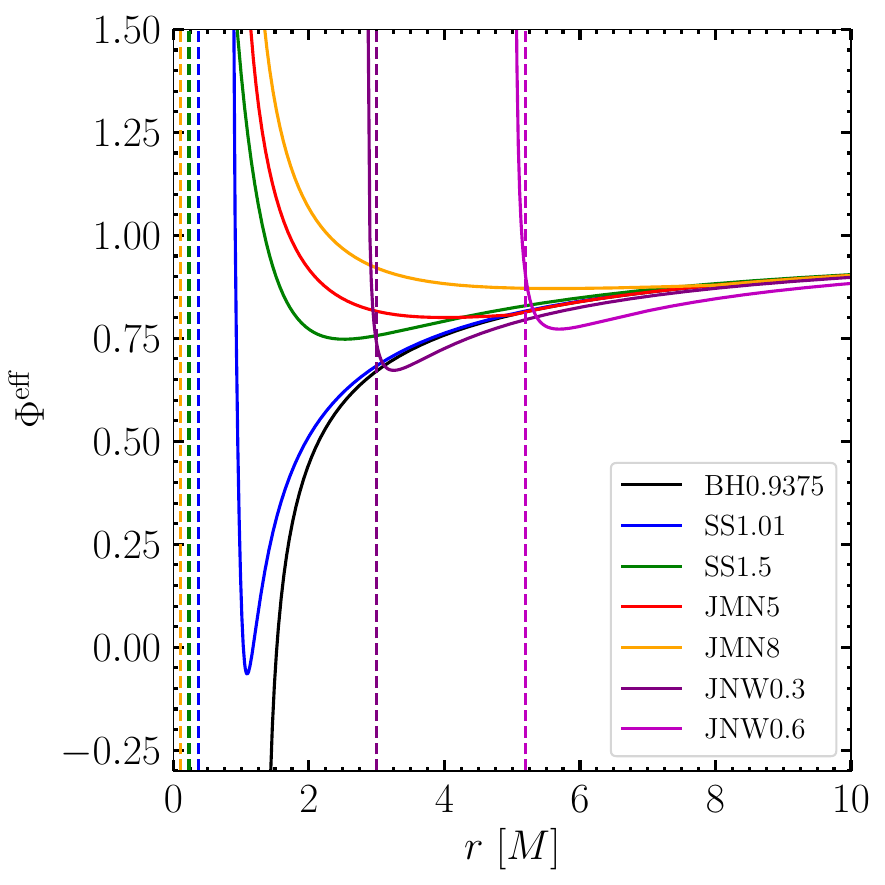}
    \caption{Effective potential in the equatorial plane for specific angular momentum $\lambda=1.65$. The NkS curves correspond to space-time models as labeled in the legend (SS1.01 through JNW0.6 from left to right). The vertical dashed lines indicate the inner edge of the computational domain for the same models (Table~\ref{IB}).}
    \label{pot_1D}
\end{figure}

We fix the inner boundary of the simulation to be just outside the location of the singularity on the equatorial plane (Table \ref{IB}). This is important to ensure that the simulation domain remains regular, corresponding to the physical space-time structure. Interestingly, all three space-times have different radial locations for the singularity without a horizon, and to understand their salient properties, we show in Figs.~\ref{pot}, \ref{pot_1D} the effective potential in the equatorial plane. 

The expression for the effective potential can be given as~\citep{Kozlowskietal1978, Dihingia:2018tlr},
\begin{equation} \label{potential_eq}
\Phi^{\rm eff}=1+\frac{1}{2}\ln{\phi},
\end{equation}
and,
\begin{equation}
\phi=\frac{g_{t\phi}^2-g_{tt}g_{\phi \phi}}{(g_{\phi \phi}+ 2 \lambda g_{t\phi} +\lambda^2 g_{tt})}.
\end{equation}
where $\lambda=-u_{\phi}/u_t$ is the specific angular momentum. 

For Fig.~\ref{pot}, we have chosen $\lambda=0$ to understand the effects of pure gravity. 
For comparison with the NkS, in the top left panel we show a Kerr black hole with spin $a_*=0.9375$. The BH potential increases monotonically from the horizon outwards, or rather, in our figure, from the inner simulation boundary outwards. This is not the case for the naked singularities, as can be seen in the next four panels. For the mildly rotating SS, the Kerr NkS with $a_*=1.01$, the effective potential has a minimum on a quasi-spherical surface (note the dark blue shell in the second panel). This is largely true also for the JMN NkS, particularly the JMN5 model. For the more rapidly spinning SS ($a_*=1.50$), there is a deeper quasi-toroidal potential well close to the equatorial plane that does not repel zero-angular momentum matter. Kerr SS gravity can only support a levitating perfect fluid shell of zero angular momentum for spins $a_*\le1.299$ \citep{Pugliese2023}. A striking feature of the JMN and SS1.50 space-times is the presence of a repulsive region at higher latitudes (seen in  Fig.~\ref{pot} as a region of higher values of the potential at low values of $\theta$ and $r$).

The JNW NkSs in our simulations are different: the last panel (bottom right)  of Fig.~\ref{pot} seems to be showing a monotonically increasing effective potential for the JNW0.3 NkS, reminiscent of a black hole.
However, this is not a physical effect, but a numerical resolution issue---because of the very rapid increase of the potential near the singularity, our simulation domain does not capture the repulsive core of the potential in this case. Because of this numerical problem, in our simulations of the JNW NkS, the absorbing boundary at the inner edge of the computational domain mimics a black hole horizon.
\begin{figure*}[t]
    \centering
  \includegraphics[width=0.9\textwidth]{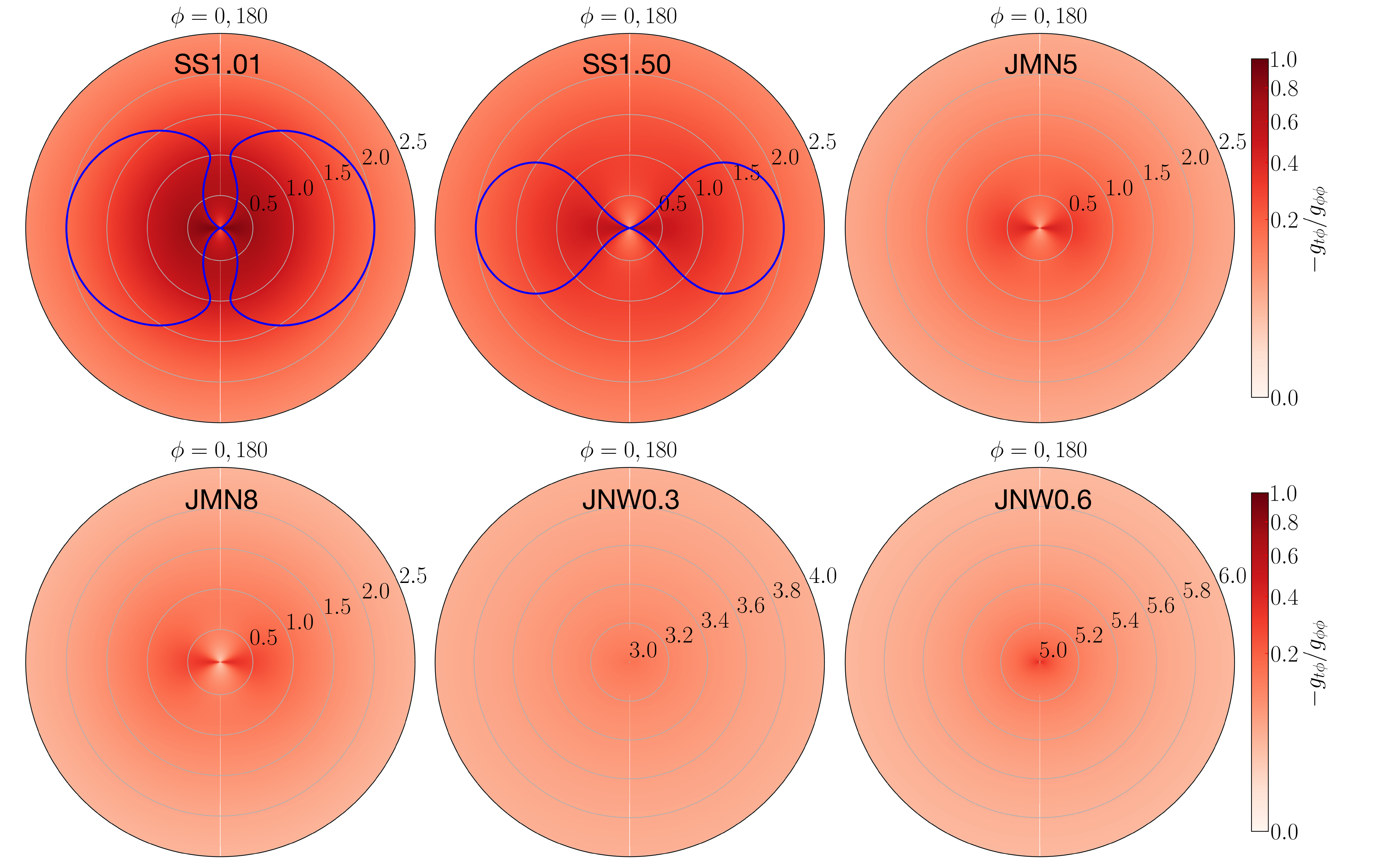}
    \caption{Ergosphere (the blue contour) and angular velocity color map in the poloidal plane for each NkS space-time.}
    \label{ergo}
\end{figure*}

In reality, the accreting matter has a non-zero specific angular momentum, and Fig.~\ref{pot_1D} exhibits the effective potential in the equatorial plane with $\lambda=1.65$ for all the metrics considered. Unlike for the black hole, whose potential curve shows a monotonic behavior, all the NkS curves show a strong centrifugal barrier (the effective potential increases with decreasing $r$). The vertical dashed lines show the location of the inner edge of our computational boundaries (with the colors corresponding to the solid curves). It is immediately apparent that the JNW centrifugal barrier is cut-off (suppressed) by our numerical boundary.


These result suggest that the nature of these singularities is repulsive in general, which is quite different from a BH's attractive nature. In this work, we try to explore extreme cases. Therefore, we specifically choose two models that can mimic the general structure of the space-time with a fixed spin for each case.

In interpreting our simulation results, it will be helpful to know whether these space-times are endowed with an ergoregion (named by analogy with the ergosphere of the Kerr BH). 
When present, it can help to understand the jet launching mechanism in these space-times through the Blandford-Znajek (BZ) mechanism \citep{Blandford:1977ds}. Similarly, its absence may indicate that the BZ mechanism for the jet launching is inoperative in these compact objects. In Fig.~\ref{ergo}, we show the ergoregion in the $\phi=0$ plane in polar coordinates for different NkS space-times. The blue contour lines ($g_{tt}<0$) indicate the existence of the ergoregion, and the color map measures the angular velocity of frame dragging ($\omega=-g_{t \phi}/g_{\phi \phi}$) for each space-time. It is interesting to observe that only superspinars have an ergoregion.
It has the topology of a torus, and its extent away from the equatorial plane decreases as the spin of the superspinar increases. The JMN and JNW space-times do not have any ergoregion, suggesting a clear indication of the absence of a BZ jet. A similar decreasing trend is also seen for the angular velocity of frame dragging in all the considered NkSs. In the case of the superspinars, the angular velocity decreases very close to the singularity with increasing spin. The rotating nature still persists in the case of JMN and JNW, but their angular velocity of frame-dragging is very low compared to the superspinars.

\section{Numerical Setup} \label{sec:3}
We perform 3D GRMHD simulations of magnetized accretion flows onto different NkSs by {\tt BHAC} code \citep{Porth:2016rfi, Olivares:2019dsc}. As an initial setup, we consider a hydrostatic equilibrium torus.
To make the comparison consistent, we first fixed the mass content within the non-Kerr Fishbone $\&$ Moncrief (FM) torus \citep{Fishbone1976RelativisticFD, Uniyal:2024sdv} and started with the Standard and Normal Evolution (SANE) model. We fixed the inner edge of the torus to be $r_{\rm in}= 10.0~r_{\rm g}$ and the density maximum at $r_{\rm max}= 20.0~r_{\rm g}$ where $r_{\rm g} \equiv GM/c^2$ and $M$ is the gravitational radius and mass of the NkS.
We consider an ideal gas equation of state with adiabatic index $\Gamma_{\rm g} = 4/3$ \citep{Rezzolla:2013dea}.
We put a weak single magnetic field loop inside the torus by fixing the initial magnetic field strength with minimum plasma-beta $\beta \equiv p_{\rm g} / p_{\rm m} =100$, where $p_{\rm g}$ and $p_{\rm m}$ are the gas and magnetic pressure, respectively. The only non-zero component of the vector potential is given by $A_{\phi}=\rho/\rho_\mathrm{max}-0.2$. For the NkS, one needs to simulate the flow structure very close to the singularity. Therefore, due to the higher computational cost, we change the inner boundary position $(r_{\rm in,edge})$ of the simulation domain for different cases, which are shown in Table \ref{IB}. Here we set the ``inflow'' boundary condition at the inner boundary for all cases~\citep{Dihingia:2024cch}. On the other hand, we set the radial outer boundary of the simulation box to be $r=2500\,r_g$ and set ``outflow'' boundary conditions.
To perform simulations, we use the static mesh refinement (SMR) with an effective resolution of $256 \times 80 \times 64$. Moreover, we perform one high-resolution case with an effective resolution of $512 \times 160 \times 128$ to compare and check consistency, which is shown in Appendix \ref{AppendixA}. 
Note that, for JNW cases, since the singularity surface is not at the coordinate center, it becomes important to choose the inner boundary ($r_{\rm in, edge}$) of the simulation domain reasonably for consistent results. Ideally, we should consider $r_{\rm in, edge}\rightarrow r_{\rm sing}$, however, this requires a very high-resolution simulation and is numerically demanding in 3D. We performed some high-resolution axisymmetric (2D) simulations with different inner boundary positions to study their impacts on the simulation results (see Appendix~\ref{AppendixC} for detail). These simulations suggest that only the properties of the jet are sensitive to the location of the inner boundary but are expected to saturate as $r_{\rm in, edge}\rightarrow r_{\rm sing}$.

\begin{table}
\centering
\caption{Table displays the model name, spin, model parameters, and radial location of the singularity (when spherically symmetric), $r_{\rm sing}$, and of the inner computational boundary ($r_{\rm in, edge}$), in units of $r_g$.  }
  \begin{tabular}{| c | c | c | c | c |}
    \hline
     Model & spin & Model Parameter & $r_{\rm sing}$ & $r_{\rm in,edge}$\\ 
    \hline
SS1.01 & $1.01$ & $a_*=1.01$  & NA  & $0.37$  \\
    SS1.01HR & $1.01$ & $a_*=1.01$   & NA   & $0.37$  \\
     SS1.50 & $1.50$ & $a_*=1.50$  & NA  & $0.23$   \\
     JMN5 & $0.9735$ & $R_b=5.0$ & $0.00$ & $0.11$  \\
     JMN8 & $0.9375$ & $R_b=8.0$ & $0.00$ & $0.11$  \\
     JNW0.3 & $0.9375$ & $\hat{\nu}=0.3$ & $2.86$ & $3.00$   \\
     JNW0.6 & $0.9375$ & $\hat{\nu}=0.6$ & $5.00$ & $5.21$ \\
    \hline
  \end{tabular}
\label{IB}
\end{table}

In all the cases, we excite the magneto-rotational instability (MRI) inside the FM torus by applying a $2\%$ random perturbation to the gas pressure. The spin is a fixed parameter $a=0.9375$ for JMN and JNW cases. The simulations evolve up to $t=10,000\, M$ to make sure that they reach a quasi-steady state. To avoid the very low density and gas pressure region, we fix the floor values of rest-mass density, $\rho_{\rm fl} = 10^{-4} r^{-2}$ and the gas pressure $p_{\rm fl}=(10^{-6}/3) r^{-2\Gamma_{\rm g}}$. Similarly, the maximum magnetization is fixed by $\sigma_{\rm max}=100$.

\section{Results} \label{sec:4}

\subsection{Accretion rate}

\begin{figure*}[t]
    \centering
    \includegraphics[width=0.6\textwidth]{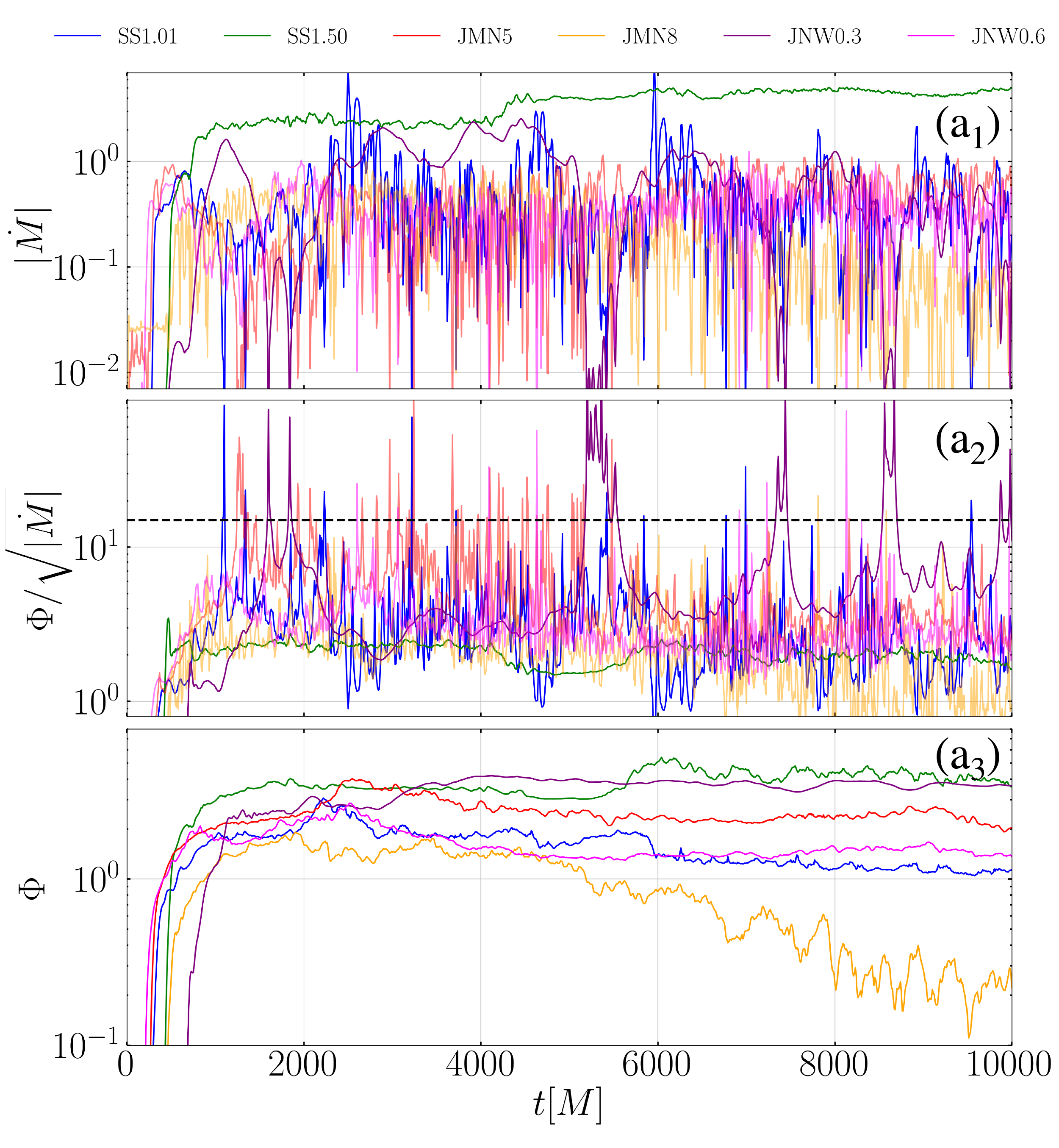}
    \caption{The volume integrated mass accretion rate ($\dot{M}$), normalized magnetic flux ($\phi / \sqrt{|\dot{M}|}$), and magnetic flux ($\phi$) for different NkS at $r= 10 \, r_g$ are shown in panels $\rm(a_1)$, $\rm (a_2)$ and $\rm (a_3)$, respectively. A black dashed horizontal line in panel $\rm (a_2)$ corresponds to the MAD limit ($\phi / \sqrt{|\dot{M}|}$=15)~\citep{Narayan:2003by, Tchekhovskoy:2012bg, McKinney:2012vh}.}
    \label{time}
\end{figure*}

Fig.~\ref{time} shows the time evolution of the mass accretion rate $(\dot{M})$, normalized magnetic flux $(\Phi / \sqrt{\dot{M}})$, and magnetic flux $(\Phi)$ for different NkS calculated at $r= 10 \, r_g$ as,
\begin{align}
 \dot{M}:=-\int_0^{2\pi}\int^\pi_0\rho u^r\sqrt{-g}\,d\theta d\phi\,,
\label{mdot}
\end{align}
and,
\begin{align}
  \Phi:=\frac{1}{2}\int_0^{2\pi}\int^\pi_0 |B^r|\sqrt{-g}d\theta d\phi\,.
\end{align}
The mass accretion rate $({\dot M})$ shows rapid fluctuations of relative amplitude unity for all of the NkS cases except SS1.50
and JNW0.3. Interestingly, SS1.50 has a stable mass accretion rate at a level several times higher than those in the other cases. JWN0.3 has a mean value of $\dot M$ comparable to the other cases, but its excursions are slower and milder than in the other fluctuating cases.
The oscillating behavior reflects the fact that the accretion rate also has negative values due to the net outflow dominating at a particular radius and time.
This oscillatory pattern also suggests that there is a possibility of either strong outflow or accumulation of matter near the singularity due to the existence of a potential well. We will study it in detail in the following sections.

These large-amplitude, rapid fluctuations are not seen in the magnetic flux $(\Phi)$ profiles. The magnetic flux keeps decreasing for JMN8 and not for JMN5. This may indicate that for JMN space-time, the matching radius $(R_b)$ has a transition point after which the space-time is not able to accumulate the magnetic flux. For other space-times, the magnetic flux saturated after $t=4000 \, t_g$. The normalized magnetic flux $(\Phi/ \sqrt{| \dot M |})$ shows a similar oscillation trend due to the accretion rate but remains in the SANE regime. A dashed horizontal line in panel $\rm (a_2)$ corresponds to the usual MAD limit ($\phi / \sqrt{|\dot{M}}|$=15)~\citep{Narayan:2003by, Tchekhovskoy:2012bg, McKinney:2012vh}.

\subsection{Outflow properties}

\begin{figure*}[t]
    \centering
    \includegraphics[width=1.0\textwidth]{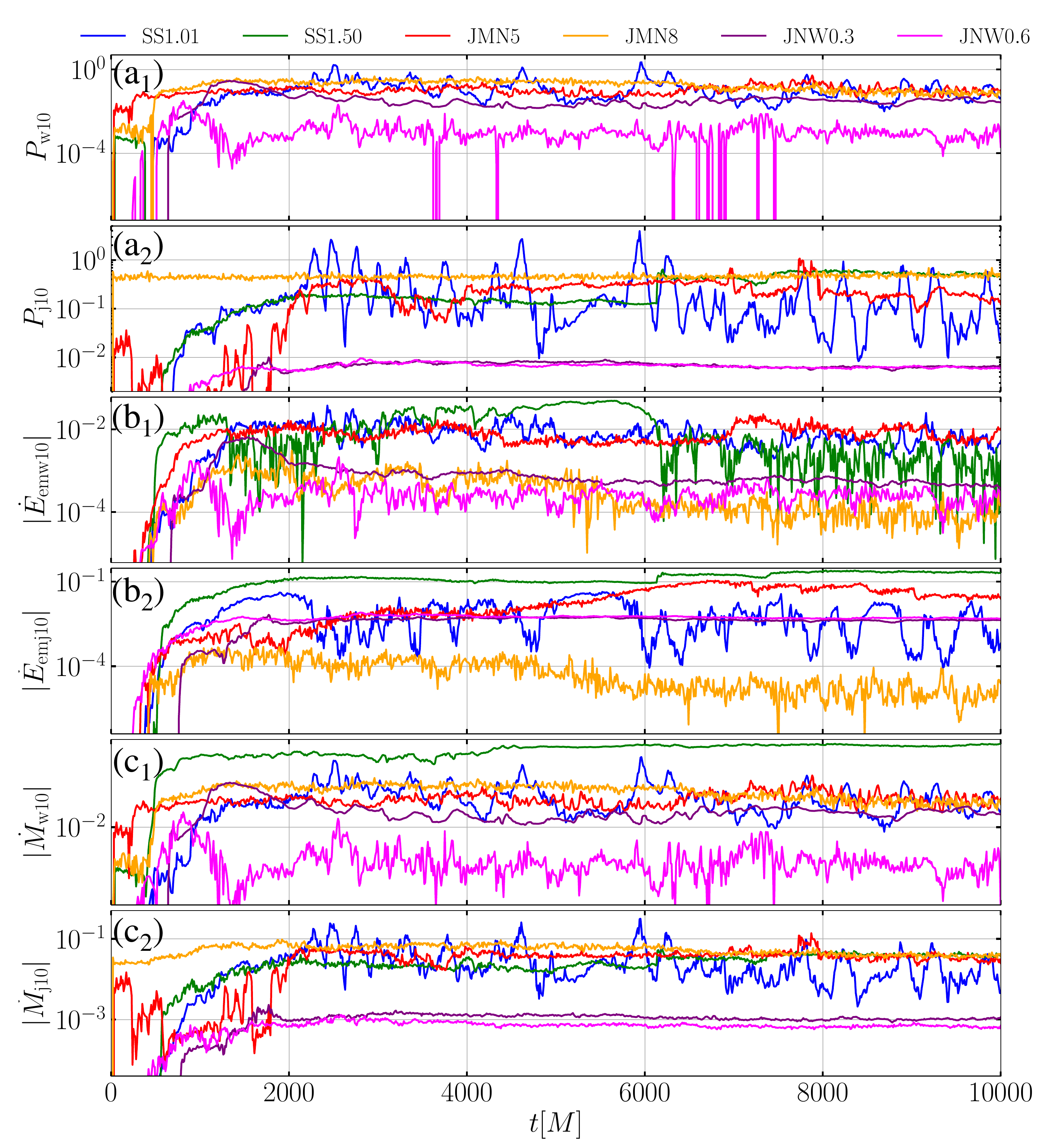}
    \caption{Time evolution of integrated outflows (jet and wind) at $r=10 \, r_g$ for different naked singularity objects. The jet and wind power are shown in (a), and their corresponding integrated electromagnetic and mass flux are shown in panels (b) and (c), respectively.}
    \label{jet}
\end{figure*}

In the previous section, we saw that the ergoregion is missing for the JMN and JNW space-time. Therefore, it is interesting to look into the jet properties of these naked singularities. In Fig.~\ref{jet}, we show the wind (panel (a$_1$)) and jet (panel (a$_2$)) time variation, the latter calculated at $r=10 \, r_g$ over the $2$-sphere with the condition where $\sigma>1$ and $-hu_t>1$ as \citep{Nathanail-etal2020},
\begin{equation}
    \begin{aligned}
  P_{\rm jet10}:=\int_0^{2\pi}\int^\pi_0 (-T^r_t -\rho u^r)\sqrt{-g}d\theta
  d\phi\,,
\label{pjet}
\end{aligned}
\end{equation}
where $T^r_t=(\rho + u_g + b^\mu b_\mu)u^ru_t-b^rb_t$ is the $(r,t)$ component of the stress-energy tensor, $\sigma=b^2/\rho$ is the magnetization, and $h$ is the specific enthalpy. Similarly, the wind power coming out of the NkS can be defined by integrating equation (\ref{pjet}) with the condition $\sigma<1$ and $-hu_t>1$.

\begin{figure*}[t]
    \centering
    \includegraphics[width=0.7\textwidth]{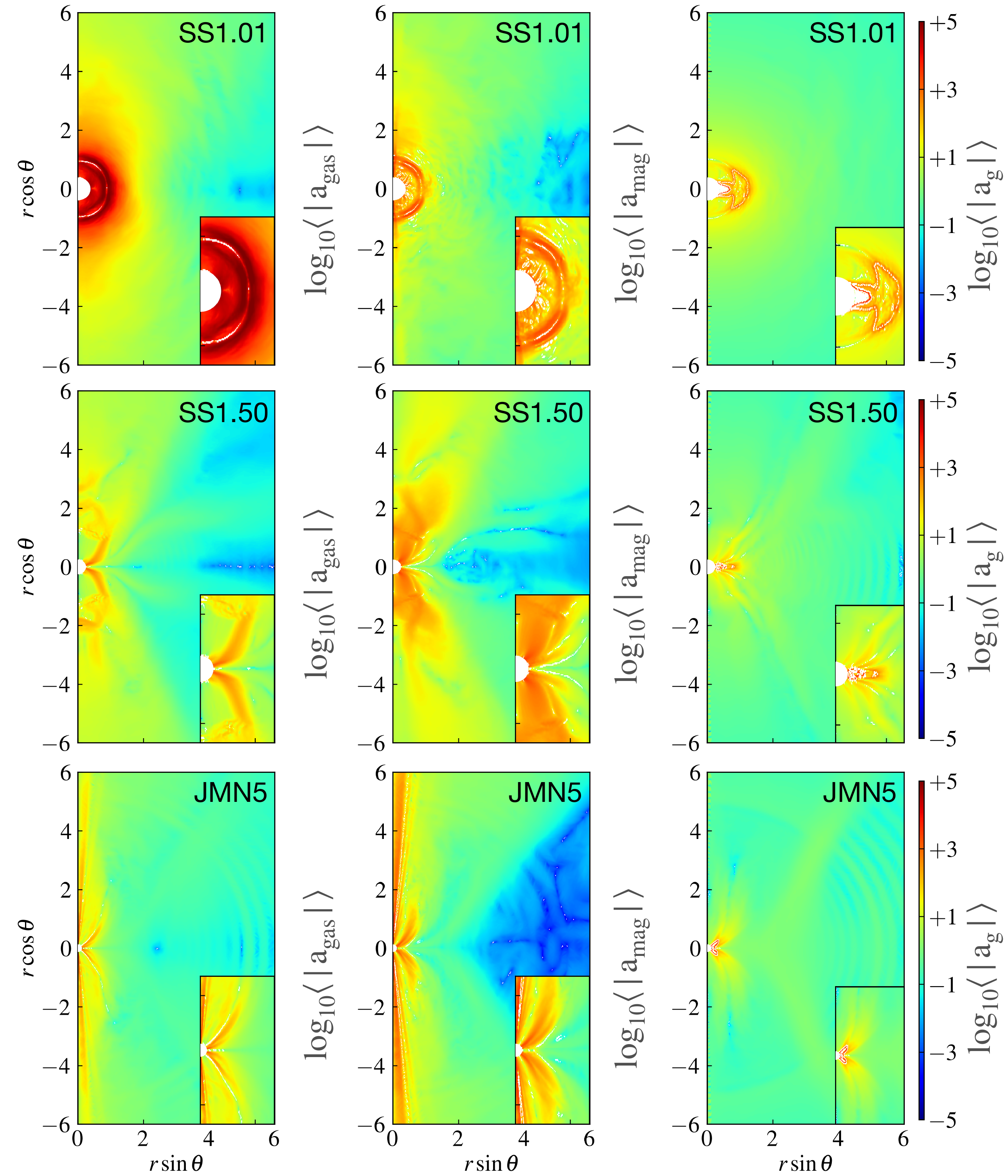}
    \caption{Time averaged ($t=8000-10000 \, t_g$) logarithmic acceleration by gas pressure $( \rm a_{gas})$, magnetic pressure $( \rm a_{mag})$ and gravity $(\rm a_{g})$ for different naked singularity objects. In each panel, the small‐scale structure is shown by a close‐up view around the singularity in the lower-right corner.}
    \label{force}
\end{figure*}

Out of all these NkS models, only the JMN and superspinar SS1.50 case show strong, steady jet power. Interestingly, as mentioned above, JMN does not exhibit any ergoregion. The two cases of the JNW have almost the same jet power, and we see oscillation for the superspinar SS1.01 case. 
It is only in SS1.01 that the (toroidal) ergoregion approaches the axis, and it is only in this case that the jet power is undergoing strong fluctuations.

Therefore, the higher, steady jet power and missing ergoregions suggest that the BZ process is not crucial to launching jets in the naked singularities. 

On the other hand, to understand the contribution of different components in the jet and wind power, we show the corresponding electromagnetic and total outward mass flux components in panels (b) and (c), respectively. The BZ process requires a strong, ordered poloidal magnetic field along the perpendicular axis of the compact objects. Such a system increases the electromagnetic component of the jet power. Previous studies suggest that the existence of the ergoregion helps to produce a strong jet through the BZ process. In the case of the superspinar, we observe that the ergoregion is present; however, the electromagnetic component is not influenced by the existence of the ergoregion alone, because SS1.50 shows a higher electromagnetic component but a smaller ergoregion (panel $(b_2)$). Similarly, JMN space-time does not have any ergoregion, but JMN5 shows a higher electromagnetic component, almost equal to that of the superspinar SS1.50. Therefore, it is not clear what mechanism is responsible for the jet power in a naked singularity. The repulsive nature of the effective potential does increase the mass flux component contributing to the jet, which is shown in the panel $(c_2)$ of Fig.~\ref{jet}. For all the NkS, the outward mass flux is either the same or higher by a factor of two than the corresponding electromagnetic component in the outflows, as can be seen by comparing panels ($c_1$) and ($b_1$), showing that a stronger reflective wall does produce stronger outward mass flux. 

In all the panels of Fig.~\ref{jet}, i.e. for all the exhibited variables (jet and wind power, electromagnetic and mass flux)  the SS1.01 NkS is unique in the level of its fluctuations. This may be related to the deep potential well, exhibited in Figs.~\ref{pot}, \ref{pot_1D}, which can serve as a reservoir of mass and energy. The more rapidly spinning SS1.50 model shows another unique feature. While at large distances, say at $r=50\,r_g$, an outflowing wind is present in all the models, close to the SS1.50 NkS the ``wind'' is inflowing: at $r=10\,r_g$ the outgoing power indicated by the corresponding (green) line in panel $(a_1)$ of Fig.~\ref{jet} rapidly decreases at early times, and remains negative (not shown) for the remainder of the simulation.
This may be related to the strongly attractive potential close to the equatorial plane in the Kerr spacetimes with $a_*\ge 1.3$, as discussed in Section~\ref{effpot}. Panel $(c_1)$ shows the large and steady mass accretion rate related to this ``inflowing wind.''

Finally, to understand the jet launching mechanism for NkS, in Fig.~\ref{force} we show vertical component of three different accelerations generated by gas pressure ($\rm \bf{a}_{gas} = -\nabla P_{gas} / \rho$), magnetic pressure ($\rm \bf{a}_{mag} = (\bf{J} \times \bf{B}) / \rho$), and gravity ($\rm \bf{a}_g = - \nabla \Phi^{eff}$). Here, $\rm P_{gas}$, $\bf{J}$, $\bf{B}$, and $\rm \Phi^{eff}$ are the gas pressure, current density, magnetic field, and effective potential, respectively. 

\begin{figure*}[t]
    \centering
  \includegraphics[width=0.9\textwidth]{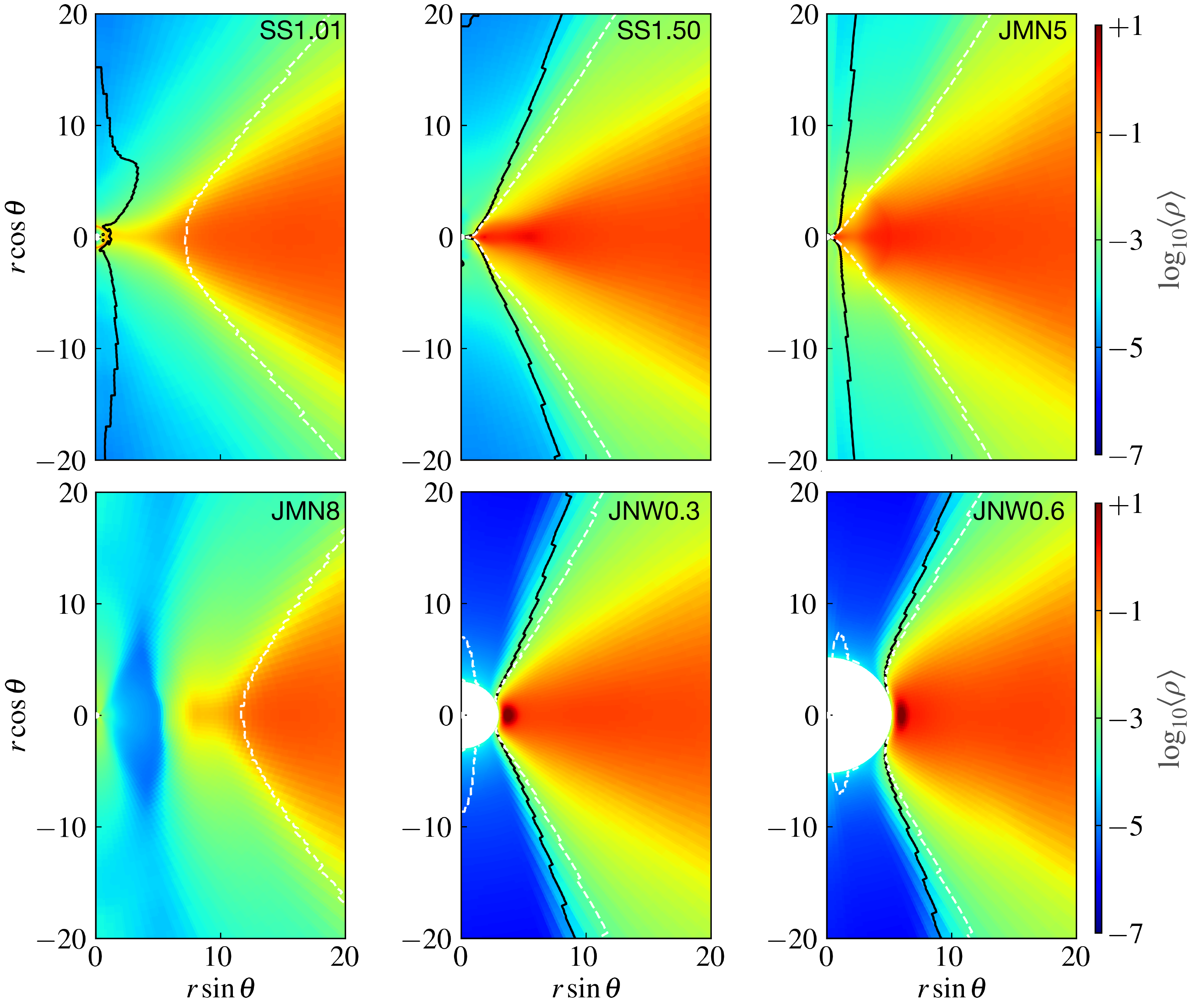}
    \caption{Azimuthal and time averaged ($t=8000-10000 \, t_g$) logarithmic density $(\rho)$ for different naked singularity objects. The black and dashed white lines represent $\sigma = 1$ and $-hu_t=1.02$ respectively.}
    \label{den}
\end{figure*}
%

In all the cases, gravitational acceleration is lower than the gradients of magnetic and gas pressure. However, due to the spherical structure of the zero-gravity surface for SS1.01, we see a clear spherical structure in all the force plots for that case. For other cases, no such spherical surface existed, but near the singularity, the higher gravity force indicates a strong repulsive force. Hence, very close to these compact objects, the gravitational force is repulsive and helps in launching the strong wind. It is important to note that the white regions within the computational domain in the figures indicate locations where the force is zero, corresponding to a zero-gravity surface for SS1.01. These points do not necessarily form a spherical surface, which is why the pattern appears different for other cases, depending on the flow angular momentum and the nature of the spacetime. More importantly, since the gravitational potential energy is being converted into thermal and magnetic energy, the effective gravitational force becomes weaker in comparison to the contributions from gas pressure and magnetic pressure. This further raises an important question, whether thermal or magnetic energy is primarily responsible for driving the outflows in the case of naked singularities. 
For superspinar SS1.01, gas pressure acceleration is much higher than magnetic pressure acceleration, which suggests the existence of the zero-gravity surface and the repulsive 
nature of gravity close to the NkS can significantly contribute to the launching mechanism of the jet and wind. However, for superspinar SS1.50, the acceleration by magnetic pressure becomes higher than the acceleration by gas pressure because here the zero gravity surface does not exist and the launching mechanism is governed by the magnetic fields twisted by ergosphere. A 3D volume rendering of the magnetization for SS1.01, presented in Appendix~A (Fig.~\ref{resolution_comp}), provides a clear illustration.
In the case of JMN5, again, acceleration by magnetic pressure becomes stronger than gas, and similar effects have been seen for the cases of JNW. 

This clarifies that the launching mechanism for the superspinars depends on the spin values. For lower spin SS1.01, it is dominated by gas pressure, and for higher spin SS1.50, magnetic pressure is more dominant, which is also the same as the case for JMN and JNW space-time. At this point, we would like to discuss the differences in the scenario from black holes. For a black hole, the effective force on the fluid element is always attractive at the horizon, but it is reflective in the case of NkS as it approaches the singularity. These differences impact the jet properties drastically (see for discussion \cite{Dihingia:2024cch}). Additionally, unlike black holes, in all our simulation cases, matter cannot reach the NkS because of the infinite potential wall at the singularity, and hence the accretion flow cannot exchange energy with the central NkS irrespective of the presence (SS1.01, SS1.50) or absence (JMN5,  JMN8, JNW0.3, JNW0.6) of the ergoregion. This is because it is located away from the singularity and therefore cannot tap the object’s rotational energy. Thus, the energy driving the jet/outflow is mainly supplied by the gravitational potential energy by pushing a fraction of matter deeper into the gravitational well. Consequently, we expect accumulation of matter around the NkS, which we discuss in the next subsection.

\subsection{Plasma distribution}

To further understand whether the different
NkS may be distinguished, we present the azimuthal and time-averaged $(t=8000-1000 \, t_g)$ distribution of logarithmic density $(\rho)$ in Fig.~\ref{den}. The black and white lines correspond to the $\sigma=1$ and $-hu_t=1.02$ contours, respectively. $\sigma > 1$ and $-hu_t\gtrsim1.02$ are typically considered as a jet funnel region. We see the existence of the zero gravity surface (a surface where fluid can be at rest, which also corresponds to the existence of the minima in the potential) for superspinar SS1.01 as also reported in \cite{Kluzniak:2024cxm, Dihingia:2024cch}, for other NkSs. Even though this surface has a more complicated shape for the superspinar with higher spin, SS1.50, some matter accumulates in the potential well near the equatorial plane. 

\begin{figure*}[t]
    \centering
  \includegraphics[width=0.9\textwidth]{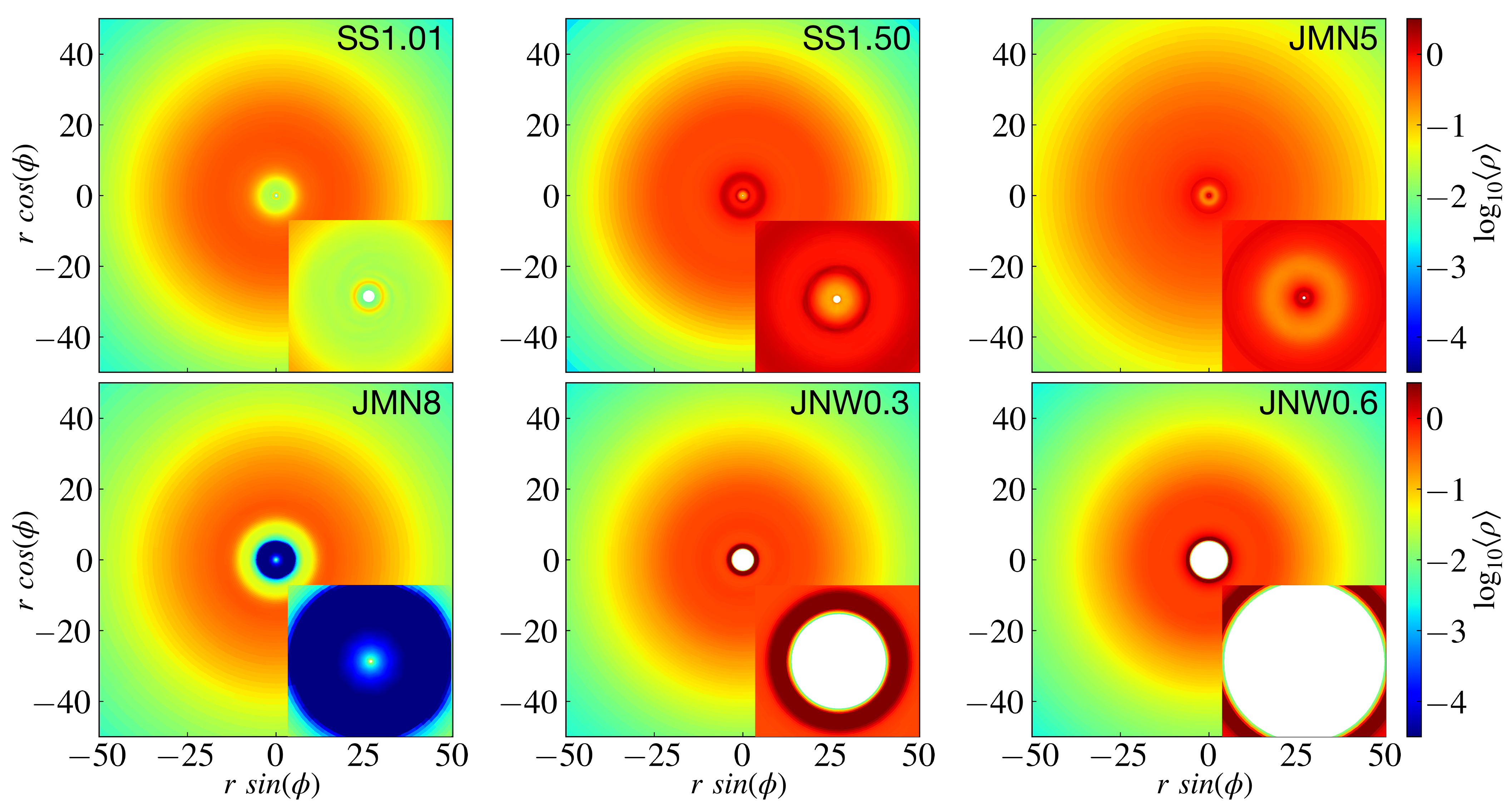}
    \caption{Time averaged ($t=8000-10000 \, t_g$) logarithmic density $(\rho)$ distribution at equatorial plane for different naked singularity objects. In each panel, the small‐scale structure is shown by a close‐up view around the singularity in the lower-right corner.}
    \label{eq_den}
\end{figure*}

The time-averaged $(t=8000-10000 \, t_g)$ logarithmic density $(\rho)$ distribution at equatorial plane is shown in Fig.~\ref{eq_den}.
Interestingly, most of the density is accumulated around the equatorial plane for all the cases.
The common feature that we can observe in the case of NkS is the accumulation of matter very close to the singularity (see \ref{eq_den}), which is not a general trend in the case of BHs (for reference, check Fig.~2 of~\cite{Dihingia:2024cch}).
For the SS1.01 superspinar, we can see the existence of a zero-gravity surface just outside of the naked singularity that is surrounded by a lower-density region. Since the fluid can remain stationary on this surface, we see the accumulation of matter there. Superspinar SS1.50 shows a similar high-density region formation very close to the NkS location, but is surrounded by a slightly higher-density region compared to the SS1.01 case. The accumulation of matter in the equatorial plane can also be seen in the case of JMN5. However, in the case of JMN8, we do not observe such a higher-density region close to the singularity. On the contrary, a very low-density region can be seen surrounded by a slightly higher-density region. This suggests a strongly reflective nature of the potential wall such that matter never reaches the singularity location.

The JNW cases do not show comparable fluid structures. Again, this suggests that our inner boundary was placed too close to capture the reflective potential wall of the JNW NKS.
Consequently, in both cases, we see a monotonically increasing density as the inner boundary is approached.

\subsection{{Reflective} nature of potential}
\begin{figure*}[t]
    \centering
  \includegraphics[width=0.9\textwidth]{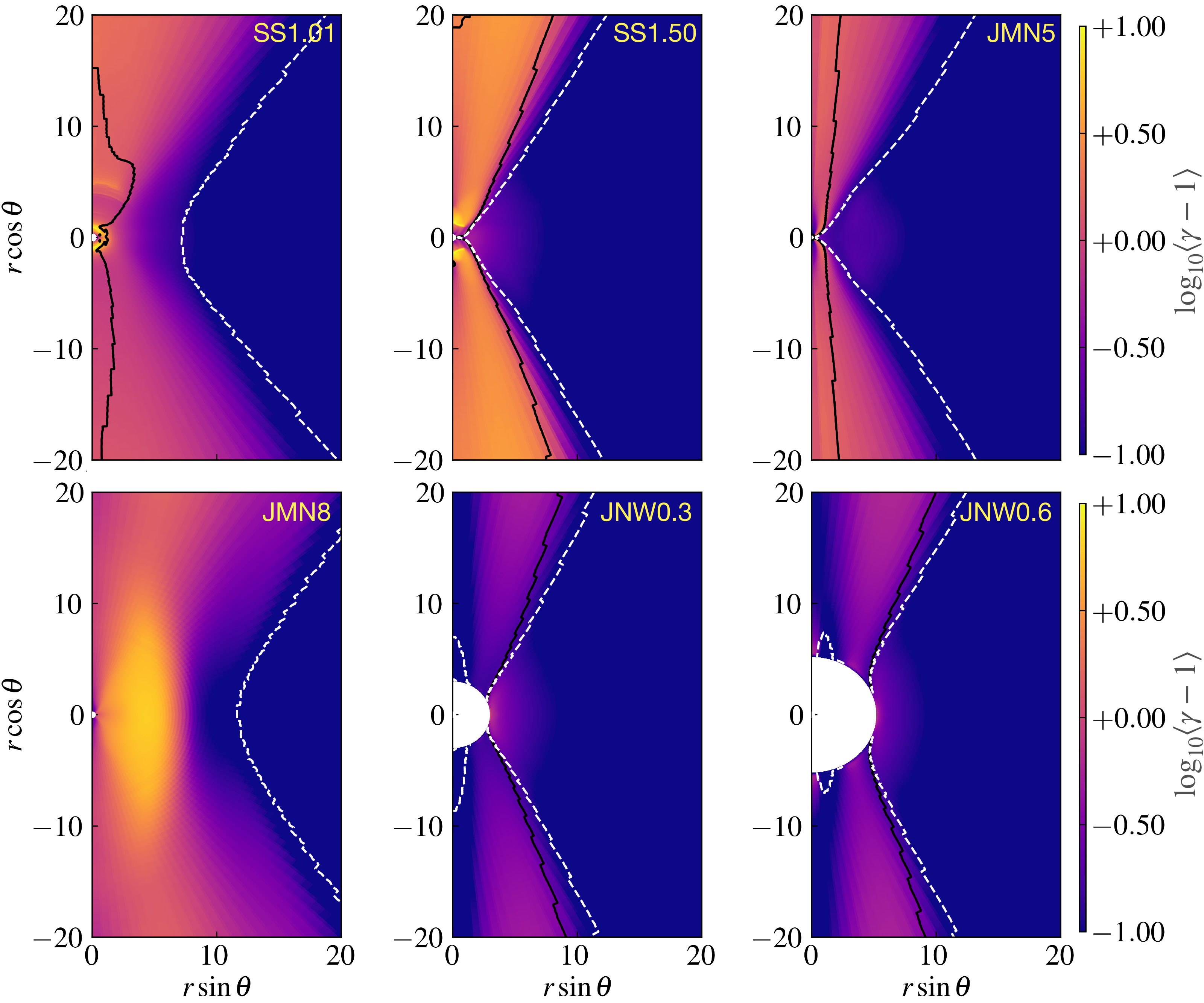}
    \caption{Same as Fig.~\ref{den} but shown in logarithmic Lorentz factor ($\gamma$).}
    \label{lfac}
\end{figure*}

To understand how strongly the repulsive nature of the potential is affecting the flow structure in polar jet and disk regions, we show the azimuthal and time-averaged $(t=8000-10000 \, t_g)$ logarithmic Lorentz factor $(\gamma)$ in Fig.~\ref{lfac}. For superspinar SS1.01, the Lorentz factor is higher at the zero-gravity surface location and decreases as we go along the polar region. This behavior is similar to superspinar SS1.50, except that the funnel region has a higher Lorentz factor compared to the SS1.01 case, suggesting strong outflow as observed in Fig.~\ref{jet}. The JMN8 case exhibits a strong reflective nature of potential, with a stronger Lorentz factor in both the funnel and the disk region compared to JMN5. Interestingly, Lorentz factors drop in the case of JNW cases, making the weaker outflow compared to the other NkS objects 
(Fig.~\ref{jet}), probably for the numerical reasons discussed above. 

A quantitative vertically integrated azimuthal and time-averaged $(t=8000-10000 \, t_g)$ 1D profile for density, Lorentz factor, magnetization, and gas temperature can be seen in Appendix~\ref{AppendixB}.

\subsection{Magnetic field configuration}

\begin{figure*}[t]
    \centering
  \includegraphics[width=0.9\textwidth]{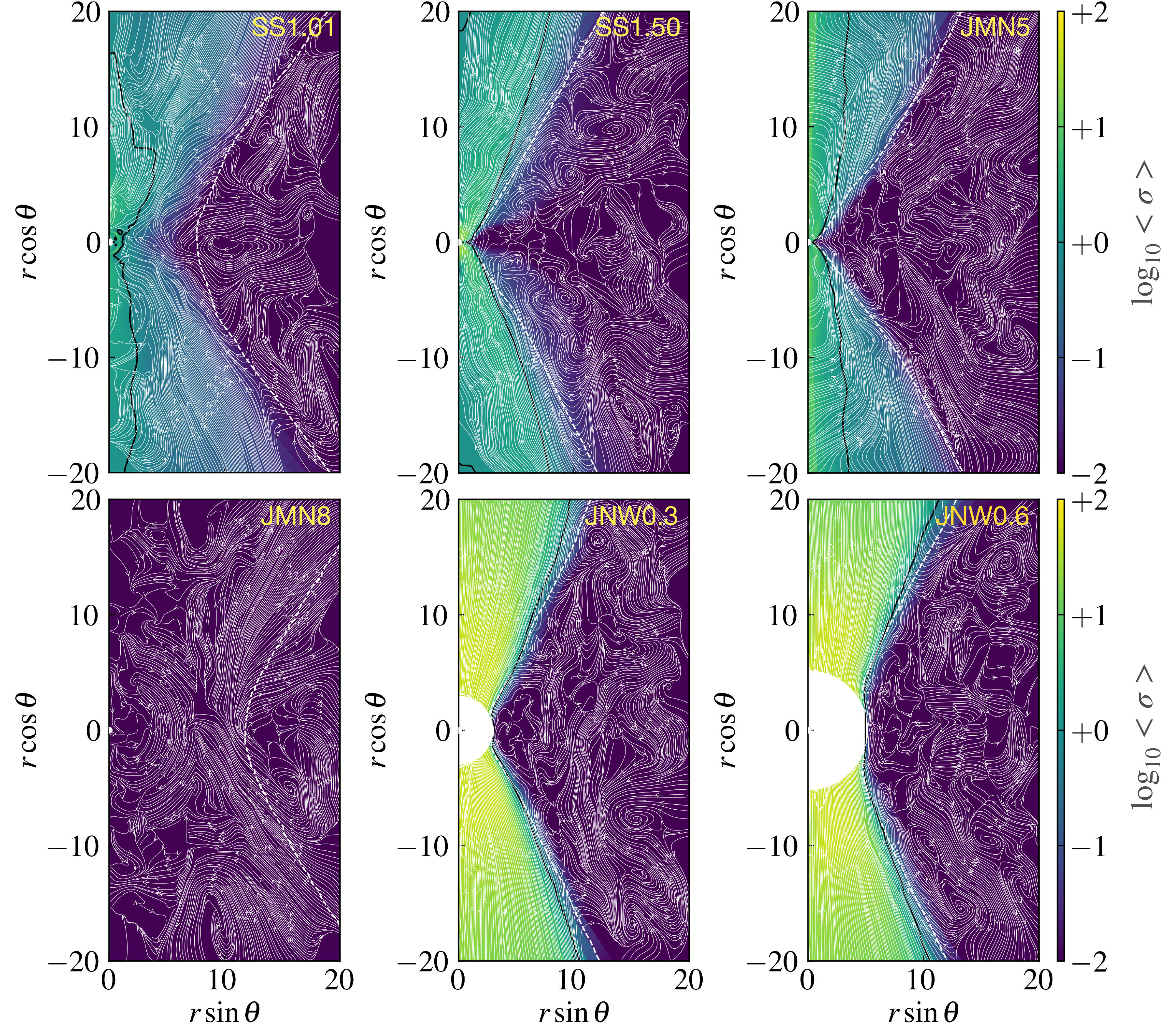}
    \caption{Same as Fig.~\ref{den} but shown in logarithmic magnetization ($\sigma$). White solid lines indicate magnetic field lines.}
    \label{sigma}
\end{figure*}

It is interesting to check how the magnetic field lines and magnetization behave in different NkS objects. In Fig.~\ref{sigma}, we show the azimuthal and time-averaged $(t=8000-10000 \, t_g)$ distribution of logarithmic magnetization $(\sigma)$. In all the cases, the magnetization is high in the polar region and low in the disk region, except for JMN8, where the magnetization is very low in the entire domain. 
In the case of JMN space-time, JMN5, we observe that the polar region has higher density compared to superspinars, making the magnetic field weaker, and eventually, for JMN8, we do not see the $\sigma\gtrsim1$ region. Such behavior of JMN8 suggests a strong reflection due to the higher effective potential wall (see Fig.~\ref{pot}). JNW space-time behaves similarly to the superspinars with a low-density region in the funnel and a larger $\sigma>1$ area. The funnel region increases for JNW0.6 due to the large singularity size compared to JNW0.3. 

For superspinar SS1.01, the magnetic field lines are slightly more turbulent compared to SS1.50, for which the field lines are more organized in the funnel region, facilitating the Blandford-Znajek (BZ) mechanism for jet propagation. In the case of superspinars, we observe some formation of magnetic islands or plasmoids, which suggests the existence of reconnection regions. They appeared in the jet region for SS1.01 and in the disk region for SS1.50. These reconnection features are a common occurrence due to the turbulent nature of the field lines. This behavior is particularly valuable for studying energy extraction in naked singularity (NkS) objects. For JMN5, the field lines are less structured compared to SS1.50, although they still follow a particular pattern in the funnel region, not originating too close to the singularity. 
This suggests the possibility of Blandford-Payne energy extraction in the JMN NkS. However, for JMN8, the reflective potential prevents the accumulation of magnetic fields or the formation of any significant structure near the singularity. In the case of JNW space-time, the magnetic fields behave similarly to those observed around BHs. 
They are more organized in the funnel region and turbulent in the disk region. In BHs, this behavior is primarily due to the BH's rotation. In the case of the JNW, this structure may arise from both the object's rotation and the reflective (repulsive) nature of the inner spacetime indicated by the significantly smaller angular frequency in JNW compared to the other cases (Fig.~\ref{ergo}); it could also reflect the numerical placement of the inner boundary (Fig.~\ref{pot_1D}).

\section{Summary and Discussion}\label{sec:5}

In this work, we investigated the dynamics of the accretion flow around various naked singularities, including superspinars, JMN, and JNW spacetimes, with the help of $3D$ GRMHD simulations. Due to the absence of an event horizon, accreting matter does not disappear from the vicinity of the NkS (however, some of it may exit the computational domain through the inner boundary of the simulation region).
The accretion flows can, in principle, extend close to the singularity, making it crucial to understand how the flow properties vary for different structures of these compact objects. To avoid irregularities and unnecessary numerical artefacts due to the divergence of the space-time at the location of the singularity, we fixed the inner boundary of the simulation right outside the location of the space-time singularity. However, for the JNW metric, for numerical reasons, this boundary was still outside the repulsive gravity region, and the resulting mass loss through the boundary weakened the outflow and the jet. To ensure a consistent comparison, we fixed the total mass content and the inner radius of the torus that feeds the accretion flow at the same location for all cases. Interestingly, all NkS spacetimes share a common feature of a reflective potential wall, though the shape, location, and properties (whether repulsive gravity or a centrifugal barrier) of this wall vary, leading to different flow properties in each case.

Figure~\ref{time}$(a_1)$ shows strong outflows are evident in the NkS for most cases, with the notable exception of SS1.50, where matter indeed accumulates, consistent with the results shown in Fig.~\ref{eq_den}. By contrast, the JMN8 case shows a weaker yet oscillatory mass accretion rate and significant outflow (see Fig.~\ref{jet}). In this scenario, most of the matter escapes outward, preventing any substantial accumulation in the immediate vicinity of the NkS (see Fig.~\ref{eq_den}). On the other hand, in cases where the gravitational well facilitates matter accumulation, such as JNW and SS1.01, we observe strong oscillatory behavior in the accretion rate, accompanied by accumulation of matter (see Fig.~\ref{eq_den}).

The jet structure is one of the most intriguing features of these compact objects, as in some cases they are capable of producing jets stronger than the BHs \citep{Kluzniak:2024cxm, Dihingia:2024cch}. Among all the NkS considered in this work, SS1.50 and JMN8 exhibit the strongest jets, followed by JMN5. To understand the formation mechanism behind these powerful jets, we analyzed the contributions of the electromagnetic and outflow mass flux components to the jet power. The electromagnetic component dominates in the case of SS1.50, while the outflow mass flux is the primary contributor for JMN8 and JMN5. As a result, the precise mechanism driving the strong jet in NkS remains unclear; however, the higher electromagnetic contribution in SS1.50 suggests that the BZ process may play a role in the jet formation for superspinars. In the case of SS1.01, oscillations in the jet power indicate turbulent behavior of the magnetic field within the funnel region. JNW space-time, on the other hand, produces a much weaker jet due to its weaker outward mass flux and electromagnetic component in our simulation (this may be an artifact of the inner boundary). The comparison between gas, magnetic, and gravity forces indicates that for superspinars, the jet-launching mechanism highly depends on the spin values. For lower spin (SS1.01), the existence of a zero-gravity surface and high gas pressure produces a strong jet compared to the magnetic pressure force. On the other hand, for the high spin (SS1.50) case, magnetic pressure dominates over gas pressure and is responsible for producing a strong jet and wind structure. In all cases, gravity is repulsive at high latitudes in the vicinity of the singularity but has a weak direct effect on the production of the jet and outflow.

Furthermore, we looked into the azimuthal and time-averaged $(t=8000-10000 \, t_g)$ flow properties like density $(\rho)$, Lorentz factor $(\gamma)$, and magnetization $(\sigma)$ to study the distinguishability between these space-times. In all cases, matter accumulates near the singularity, and for SS1.01, we observe a distinct spherical shell of matter surrounded by a low-density region. This matter accumulation is a characteristic feature that differentiates these objects from BHs, although it does not serve to distinguish NkSs from one another, apart from the very strong inflow at $r=10\,r_g$ in the SS1.50 model, as well as the strong fluctuation of power and outgoing fluxes in SS1.01, the minimally spinning Kerr NkS (both described in the discussion of Fig.~\ref{jet}). 

The equatorial profile of accreting matter shows that the density is significantly higher for JNW, JMN5, and SS1.50, while JMN8 exhibits almost no density near the singularity. A similar trend is observed for the Lorentz factor, which is higher at the zero-gravity surface for SS1.50 and in the funnel region for SS1.50 and JMN5. In the case of JMN8, a high Lorentz factor is observed in the disk region, indicating a strong outflow, as seen in panel $(c_2)$ of Fig.~\ref{jet}. JNW space-time shows a lower Lorentz factor but maintains a higher value in the funnel region. 

The magnetization $(\sigma)$ is notably higher for JNW space-time in the funnel region, suggesting the presence of strong magnetic fields. JMN8, on the other hand, shows no significant high magnetization and turbulent magnetic field structure, implying that only mass flux contributes to the strong jet formation in this case. SS1.50 and JMN5 exhibit similar magnetization in the funnel region, although the magnetic field lines are slightly less ordered. SS1.01 shows higher magnetization at the zero-gravity surface, with a turbulent magnetic field in the funnel region, leading to oscillations in the jet power.

\section{Conclusion}\label{sec:6}

In this work, we have provided a general overview of how accretion flow varies across different NkS space-times. Furthermore, our study can be extended to other NkS space-times, it is specifically focused on NkSs within the framework of general relativity. In general, our simulations reveal two robust features that appear across the different classes of NkS spacetimes. First, we consistently observe the accumulation of matter near the NkS, arising from the presence of a deep potential well that prevents material from being easily advected inward. This phenomenon can also vary with the non-existence of the zero-gravity surface, as seen in JMN8. 
Second, we find the emergence of strong outflows or jets, whose driving mechanisms can vary depending on the specific spacetime and location: gas pressure and magnetic pressure may each play a dominant role under different conditions. However, the driving energy for these outflows or jets comes from the cost of gravitational potential energy of the accumulated matter in the deep potential well for all the cases. 
Similar outflows were found for NkS accretion in the non-rotating Reissner-Nordstr\"om space-time \citep{Kluzniak:2024cxm}. A study of the figures of equilibrium in that space-time revealed that fluid close to the surface of the torus feeding the accretion flow is just barely bound. If it absorbs a little energy from the flow it may become unbound and will be free to move on an unbounded equipotential surface. For the Reissner-Nordstr\"om NkS the equipotential surfaces of non-negative energies have the shape of a funnel around the $z$ axis \citep[e.g., Fig. 11 of][]{Mishra:2024bpl}. Thus axial outflows seem to be a natural outcome for accretion flows around NkS.
Together, these results suggest that matter buildup near the singularity and the production of powerful outflows are generic characteristics of accretion flows onto NkSs.

Some studies have found that non-rotating, optically thick, spherical shells may form around non-spinning NkSs~\citep{Vieira:2023, Broderick:2024vjp}. Such quasi-spherical shells may also form around the Kerr NkS for $a_*\le 1.299$~\citep{Pugliese2023}. Studies of rotating figures of equilibrium around non-rotating NkS reveal that rotating shells will have the topology of a torus, and thus may not cover the NkS completely~\citep{Mishra:2024bpl}. In general, such shells may have a thermally emitting photosphere, and may obstruct initially ingoing null geodesics, thereby making tests based on directly observing lensed emission inside the shadow more challenging~\citep{Broderick:2024vjp}.

Additionally, NkSs are considered strong mimickers of BH shadows, and therefore, in a subsequent paper, we will explore the radiative properties of these objects based on the simulations performed here. The current study highlights the significant dynamical differences between NkSs and BHs, yet a comprehensive understanding of their radiative properties remains essential for further characterization.

\begin{acknowledgments}
The authors would like to thank the anonymous reviewer for their
constructive comments to improve the manuscript content. This research is supported by the National Key R\&D Program of China (Grant No.\,2023YFE0101200), the National Natural Science Foundation of China (Grant No.\,12273022), the Research Fund for Excellent International PhD Students (grant No. W2442004) and the Shanghai Municipality orientation program of Basic Research for International Scientists (Grant No.\,22JC1410600). I.K.D. acknowledges the TDLI postdoctoral fellowship for financial support. The simulations were performed on TDLI-Astro cluster and Siyuan Mark-I at Shanghai Jiao Tong University.
\end{acknowledgments}

\appendix

\section{Resolution test}\label{AppendixA}

\begin{figure*}[t]
    \centering
  \includegraphics[width=0.48\textwidth]{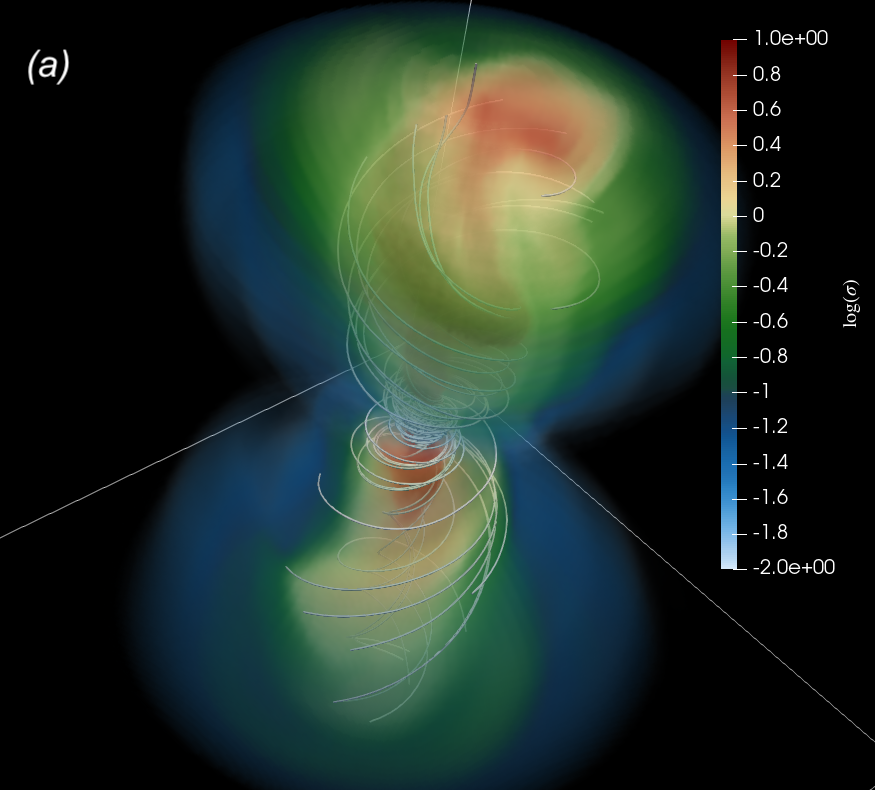}
  \includegraphics[width=0.48\textwidth]{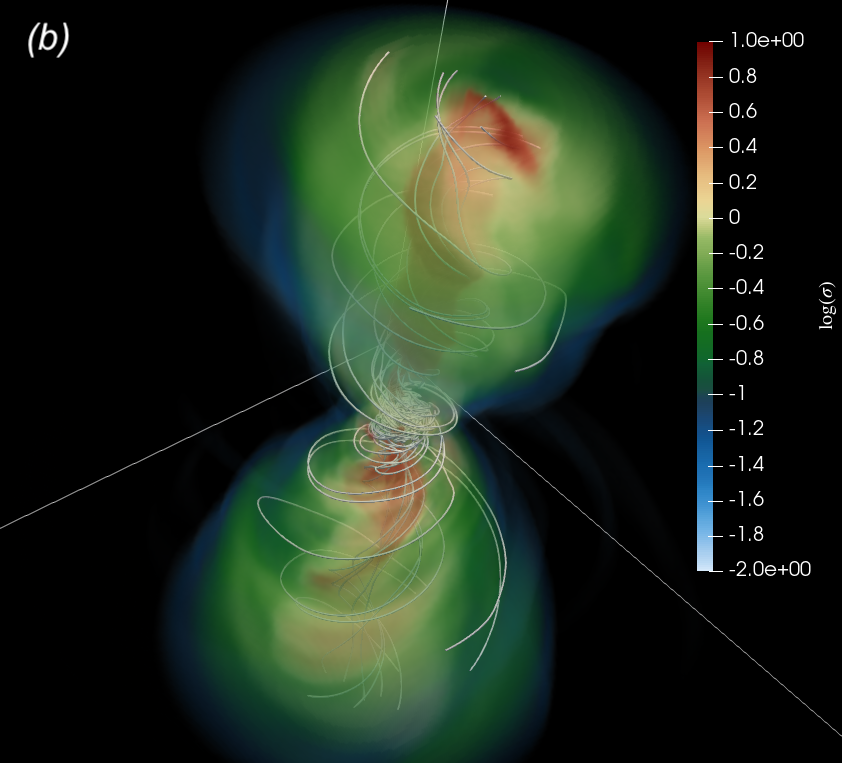}
    \caption{3D volume rendering of magnetization ($\sigma$) for (a) fiducial ($256\times80\times64$) and (b) high-resolution ($512\times160\times128$) SS1.01 cases within radii $r=20\,r_g$ at simulation time $t=5000\,t_g$. The solid tubes correspond to the magnetic field lines. }
    \label{resolution_comp}
\end{figure*}

In this section, we compare our fiducial run with the high-resolution run for a consistency test. We perform it only for the superspinar case SS1.01. A similar test is shown in \cite{Dihingia:2024cch} by comparing the density distribution and accretion rate in Appendix D of the reference. These comparisons suggest that the qualitative features around NkS remain unaltered with these choices of resolutions. Therefore, in this section, we add additional comparisons of magnetization ($\sigma$) and magnetic field lines in Fig.~\ref{resolution_comp}. This figure suggests that the jet morphology is quite similar for both resolutions. Whereas, the jet is slightly thinner for the high-resolution case than for the low-resolution case. Moreover, the qualitative behavior of the magnetic fields looks very similar. Overall, the differences in magnetization and magnetic field structures between the two simulations with different resolutions are minimal, indicating that the jet morphology is largely preserved regardless of resolution choice. This consistency reinforces the robustness of our findings around NkS. 

\section{Accretion flow properties}\label{AppendixB}

\begin{figure}[t]
    \centering
  \includegraphics[width=0.5\textwidth]{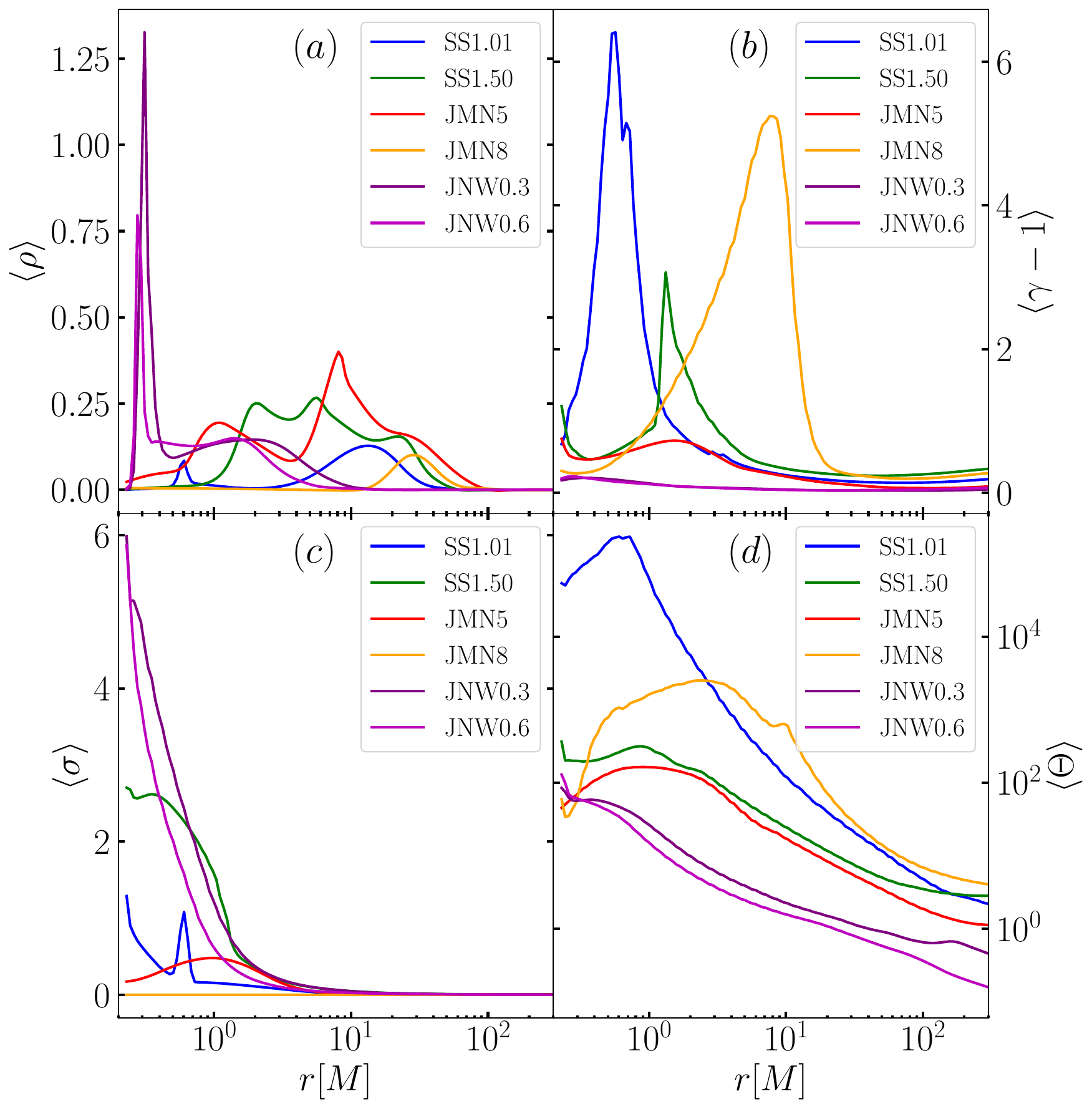}
    \caption{Time-averaged ($t=8000-10000 \, t_g$) radial profiles for density ($\rho$), lorentz factor ($\gamma -1$), magnetization $(\sigma)$, and gas temperature ($\theta$) for different naked singularity objects.}
    \label{radial}
\end{figure}

To analyze the quantitative nature of the flow properties for different NkSs, in Fig.~\ref{radial} we calculated the vertically integrated azimuthal and time-averaged mean $(a)$ density $(\langle \rho \rangle)$, $(b)$ Lorentz factor $( \langle \gamma -1 \rangle)$, $(c)$ magnetization $( \langle \sigma \rangle)$, and $(d)$ gas temperature $(\langle \Theta = P_g/\rho \rangle)$ respectively. We calculated the quantities $q(r, \theta, \phi, t)$ first as shell averaged:
\begin{equation}
\langle {\text{q}}(r, t) \rangle := \frac{\int_0^{2\pi} \int_{\theta_{\min}}^{\theta_{\max}} \sqrt{-g} \, q(r, \theta, \phi, t) \, d\phi \, d\theta}{\int_0^{2\pi} \int_{\theta_{\min}}^{\theta_{\max}} \sqrt{-g} \, d\phi \, d\theta}
\vspace{1em}
\end{equation}
which is later time-averaged over $t=8000-10000 \, t_g$. As shown in Fig.~\ref{den}, the density (panel $(a)$) exhibits a peak in all cases before reaching the singularity, indicating the accumulation of matter in the equatorial plane. For JNW space-time, this peak is sharp and higher, followed by an instant decay before reaching the singularity. JMN5 and superspinar show similar density variations, while JMN8 displays the lowest density, suggesting a strong reflective wall, as discussed in previous sections.
On the other hand, the Lorentz factor (panel $(b)$) is significantly higher for SS1.01, followed by JMN5 and SS1.50, compared to the other cases near the singularity. This implies stronger outflows in these objects, as observed in Figs.~\ref{lfac} and \ref{jet}. JNW space-time has a lower Lorentz factor, resulting in a reduced mass flux. The accumulation of both matter and magnetic field is influenced by the space-time structure, and as such, density and magnetization exhibit similar trends. Consequently, magnetization (panel $(c)$) is notably higher for JNW space-time compared to the other cases, indicating a strong possibility of the BZ process contributing to the jet power. However, it remains unclear whether the BZ process is the sole mechanism responsible for jet formation in NkS objects. For JMN8, the magnetization is weaker due to the strong repulsive nature of the spacetime, and yet we see a strong jet outflow (see Fig.~\ref{jet}). Finally, we examine the variation in the gas temperature (panel $(d)$) for all cases. SS1.01 shows the highest temperature, followed by JMN5. Despite JNW space-time exhibiting higher density and magnetization, its Lorentz factor and gas temperature are lower, suggesting that the larger singularity suppresses both heating and outflow from the compact object.

\section{Inner boundary location comparison for JNW}\label{AppendixC}
As discussed in the main text, the JNW results can be sensitive to the choice of the simulation’s inner boundary position. Because the reflective wall lies very close to the singularity (see Fig.~\ref{pot_1D} and Table~\ref{IB}), placing the inner boundary arbitrarily near the singularity is computationally demanding. In 2D, however, due to the less computational cost, we can use local mesh refinement to increase the number of cells near the diverging singularity and thereby avoid the singularity despite setting the inner boundary very close. Accordingly, we perform a set of 2D simulations considering different inner-boundary locations (see Table~\ref{JNW_IB}) with an effective resolution $1024 \times 512$ (with two levels, where maximum resolution is set for $r<=60\,r_g$). This helps us understand the possible impacts of inner boundary locations on the simulation results.

\begin{figure}[t]
    \centering
  \includegraphics[width=0.5\textwidth]{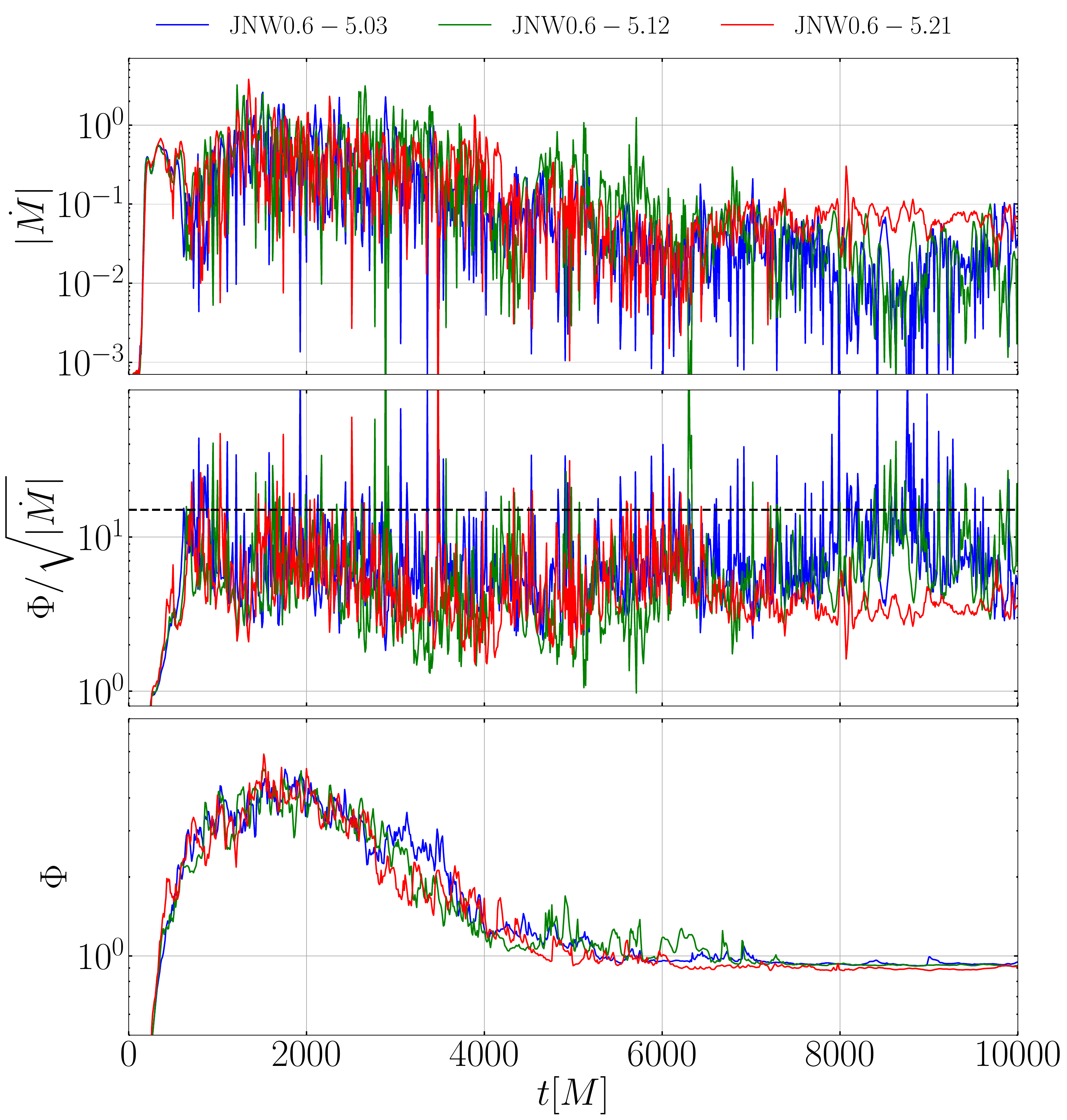}
    \caption{The surface integrated mass accretion rate ($\dot{M}$), normalized magnetic flux ($\phi / \sqrt{|\dot{M}|}$), and magnetic flux ($\phi$) for JNW0.6 NkS at $r= 10 \, r_g$ for different inner boundary location (see Table \ref{JNW_IB}).}
    \label{JNW_2D_flux}
\end{figure}

\begin{table}
\centering
\caption{Table displays the model name, spin, model parameters, and radial location of the singularity (when spherically symmetric), $r_{\rm sing}$, and of the inner computational boundary ($r_{\rm in, edge}$), in units of $r_g$ for 2D JNW simulation.  }
  \begin{tabular}{| c | c | c | c | c |}
    \hline
     Model & spin & Model Parameter & $r_{\rm sing}$ & $r_{\rm in,edge}$\\ 
    \hline
     JNW0.6-5.02 & $0.9375$ & $\hat{\nu}=0.6$ & $5.00$ & $5.02$ \\
     JNW0.6-5.03 & $0.9375$ & $\hat{\nu}=0.6$ & $5.00$ & $5.03$ \\
     JNW0.6-5.12 & $0.9375$ & $\hat{\nu}=0.6$ & $5.00$ & $5.12$   \\
     JNW0.6-5.21 & $0.9375$ & $\hat{\nu}=0.6$ & $5.00$ & $5.21$ \\
    \hline
  \end{tabular}
\label{JNW_IB}
\end{table}
Figure~\ref{JNW_2D_flux} presents the surface-integrated mass accretion rate $\dot{M}$, the normalized magnetic flux $\phi/\sqrt{|\dot{M}|}$, and the magnetic flux $\phi$ for JNW0.6 NkS evaluated at $r=10\,r_g$. All runs exhibit qualitatively similar mass accretion rates and magnetic flux.
Towards the end of the simulation, the value of the accretion rate is lower for a closer inner boundary. This suggests that when accretion flow has access to very close to the singularity, a stronger outflow is expected due to the reflective nature of the potential in that region. Due to that, the normalized magnetic flux also has a higher value for a closer inner boundary.

\begin{figure}[t]
    \centering
  \includegraphics[width=0.45\textwidth]{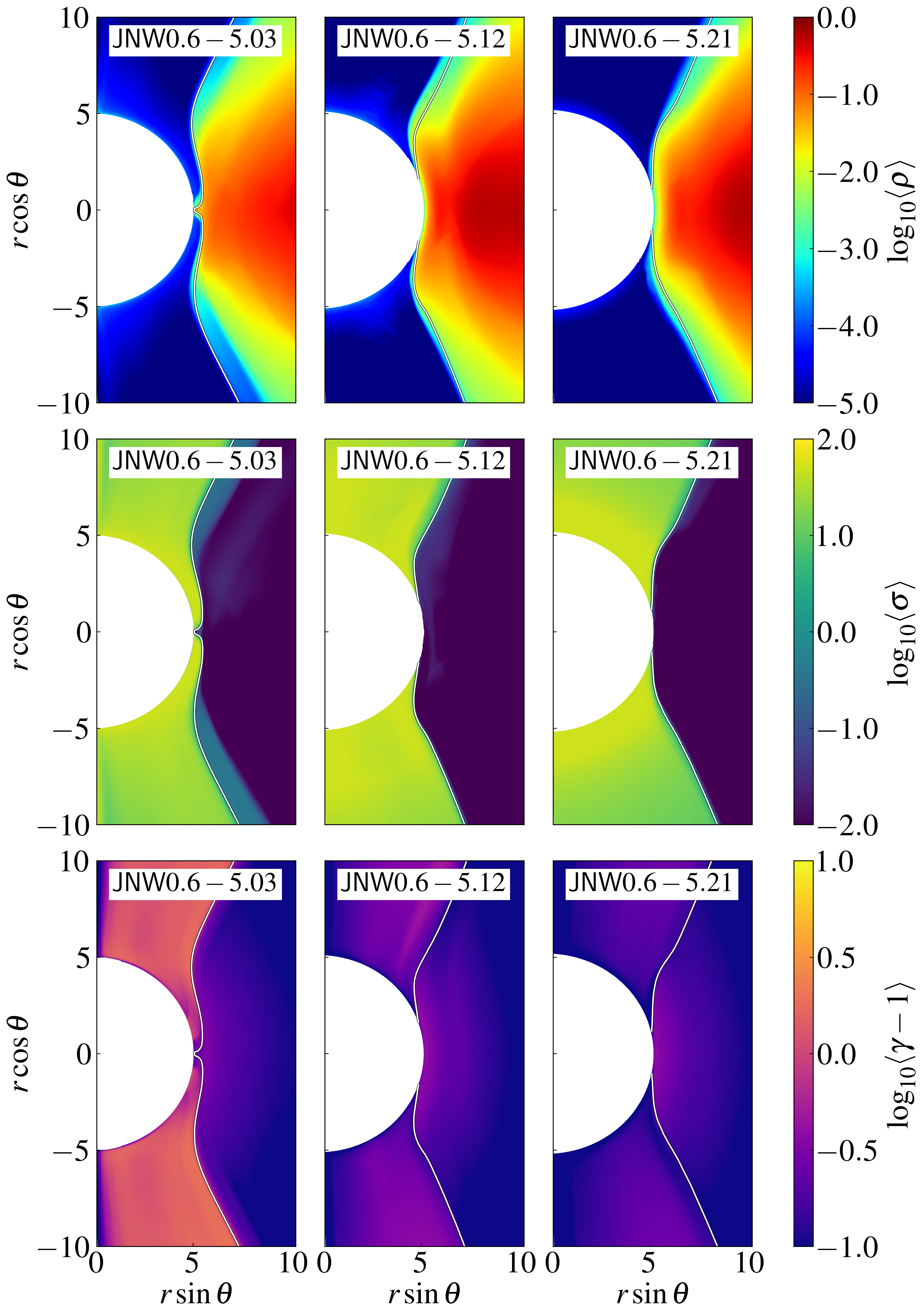}
    \caption{Time-averaged ($t=8000-10000 \, t_g$) logarithmic density $(\rho)$, magnetization ($\sigma$), and Lorentz factor ($\gamma$) for JNW0.6 for different inner boundary location (see Table \ref{JNW_IB}). The white lines represent $\sigma = 1$.}
    \label{JNW_2d}
\end{figure}

\begin{figure}[t]
    \centering
  \includegraphics[width=0.5\textwidth]{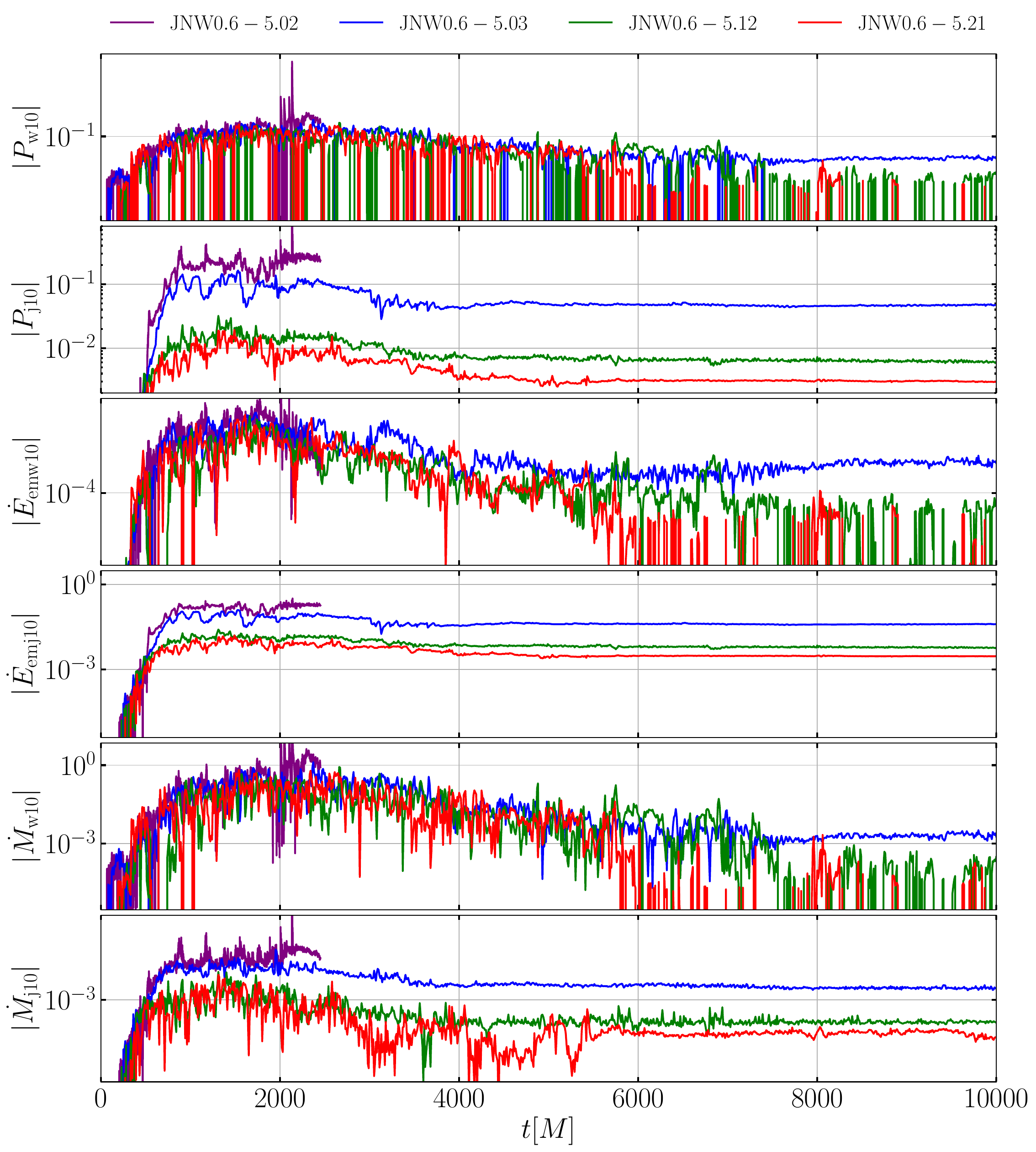}
    \caption{Time evolution of integrated outflows (jet and wind) at $r=10 \, r_g$ for JNW0.6 naked singularity for different inner boundary location (see Table~\ref{JNW_IB}).}
    \label{JNW_2d_jet}
\end{figure}

Furthermore, Figure~\ref{JNW_2d} shows logarithmic density $(\rho)$, magnetization ($\sigma$), and Lorentz factor ($\gamma$) for different inner boundary locations. In the case JNW0.6-5.03, we find that a very small amount of matter can approach the singularity very closely in the equatorial region, whereas the density structure of the reflective surface above the equator is comparable across all cases. 
Despite that, the absolute value of density at the inner boundary on the equatorial plane for the case JNW0.6-5.03 is very small $\rho_{\rm in}\sim 4.8 \times 10^{-4}$ in code units. If we increase the resolution further and the inner boundary is considered as $r_{\rm in, edge}\rightarrow r_{\rm sing}$, the value of $\rho_{\rm in}$ is expected to become zero. This is because, at the height of potential is infinite for flow with any non-zero angular momentum.
The spatial extent of highly magnetized zones is essentially unchanged, as indicated by the $\sigma=1.0$ contour; however, the Lorentz factor rises since flow could access a more reflective region closer to the singularity, which in turn increases the jet power (see Fig.~\ref{JNW_2d_jet}). Additionally, wind-related quantities show little variation between all cases, but jet power is amplified when the inner boundary is moved inward (JNW0.6-5.03). Nonetheless, if we take the inner boundary even closer to the singularity by increasing the effective resolution $2048 \times 512$ for the case JNW0.6-5.02 shown in Fig.~\ref{JNW_2d_jet}, the jet power seems to be similar to JNW0.6-5.03, providing an insight that the matter will face a similar strong reflective wall and fluid will not be able to reach the singularity.
However, it remains to be determined whether this enhancement will be the same in fully 3D simulations, although some influence on jet power is anticipated. In summary, it is noted that the bulk dynamics are largely independent of the inner-boundary placement for the JNW configurations explored here (if it is set reasonably close to the singularity), but inward boundary positions can produce higher Lorentz factors and a corresponding increase in jet power, which is also expected to saturate as $r_{\rm in, edge}\rightarrow r_{\rm sing}$.

\bibliography{sample631}{}
\bibliographystyle{aasjournal}

\end{document}